\documentclass[pdftex]{pasj01}

\Received{}
\Accepted{}
 
\usepackage{newtxtext}
\usepackage{txfonts} 
\usepackage[T1]{fontenc}
\usepackage{ae,aecompl}
\usepackage{multirow}
\usepackage{threeparttable}
\usepackage{lscape}
\usepackage[switch,mathlines]{lineno}

\graphicspath{{New/}}
\let\biblio\bibliography
\let\bibsty\bibliographystyle
\renewcommand{\bibliography}[1]{\expandafter\biblio{New/#1}}
\renewcommand{\bibliographystyle}[1]{\expandafter\bibsty{New/#1}}

\newcommand{\cplus}{[C\,{\sc ii}]}

\def\kms    {km\,s$^{-1}$}
\def\Lsun    {L$_\odot$}

\begin{document} 

\title{ALMA Deep Field in SSA22: Reconstructed \cplus\ Luminosity Function at $z$ = 6}

\author{Natsuki\,H.~\textsc{Hayatsu}\altaffilmark{1}}
\altaffiltext{1}{Waseda Research Institute for Science and Engineering, 3-4-1 Okubo, Shinjuku, Tokyo 169-8555, Japan}
\email{hayatsu@aoni.waseda.jp}

\author{Rob\,J.~\textsc{Ivison}\altaffilmark{2}}
\altaffiltext{2}{European Southern Observatory, Karl-Schwarzschild-Strasse~2, D-85748 Garching, Germany}

\author{Paola~\textsc{Andreani}\altaffilmark{2}}

\author{Fabrizia~\textsc{Guglielmetti}\altaffilmark{2}}

\author{Zhi-Yu~\textsc{Zhang}\altaffilmark{3,4}}
\altaffiltext{3}{School of Astronomy and Space Science, Nanjing
  University, Nanjing 210023, People's Republic of China}
\altaffiltext{4}{Key Laboratory of Modern Astronomy and Astrophysics (Nanjing University), Ministry of Education, Nanjing 210093, People's Republic of China}

\author{Andy~\textsc{Biggs}\altaffilmark{5}}
\altaffiltext{5}{UK Astronomy Technology Centre, Royal Observatory, Blackford Hill, Edinburgh EH9 3HJ}

\author{Hideki~\textsc{Umehata}\altaffilmark{6,7}}
\altaffiltext{6}{Institute for Advanced Research, Nagoya University, Furocho, Chikusa, Nagoya 464-8602, Japan}
\altaffiltext{7}{Department of Physics, Graduate School of Science, Nagoya University, Furocho, Chikusa, Nagoya 464-8602, Japan}

\author{Yuichi~\textsc{Matsuda}\altaffilmark{8,9}}
\altaffiltext{8}{Graduate University for Advanced Studies (SOKENDAI), Osawa 2-21-1, Mitaka, Tokyo 181-8588, Japan}
\altaffiltext{9}{National Observatory of Japan, 2-21-1 Osawa, Mitaka, Tokyo 181-8588, Japan}

\author{Naoki~\textsc{Yoshida}\altaffilmark{10,11}}
\altaffiltext{10}{Department of Physics, The University of Tokyo, 7-3-1 Hongo, Bunkyo, Tokyo 113-0033, Japan}
\altaffiltext{11}{Kavli Institute for the Physics and Mathematics of the Universe (WPI),Todai Institutes for Advanced Study, The University of Tokyo, Kashiwa, Chiba 277-8583, Japan}

\author{Mark~A.~\textsc{Swinbank}\altaffilmark{12}} 
\altaffiltext{12}{Centre for Extragalactic Astronomy, Department of Physics, Durham University, South Road, Durham, DH1 3LE UK}

\author{Kotaro~\textsc{Kohno}\altaffilmark{13,14}}
\altaffiltext{13}{Department of Astronomy, Graduate school of Science, The University of Tokyo, 7-3-1 Hongo, Bunkyo-ku, Tokyo 133-0033, Japan}
\altaffiltext{14}{Research Center for the Early Universe, The University of Tokyo, 7-3-1 Hongo, Bunkyo, Tokyo 113-0033, Japan}

\author{Yoichi~\textsc{Tamura}\altaffilmark{15}} 
\altaffiltext{15}{Department of Physics, Graduate School of Science, Nagoya University, Furocho, Chikusa, Nagoya, Aichi 464-8602, Japan}

\author{Bunyo~\textsc{Hatsukade}\altaffilmark{16}} 
\altaffiltext{16}{Institute of Astronomy, Graduate School of Science, The University of Tokyo, 2-21-1 Osawa, Mitaka, Tokyo 181-0015, Japan}

\author{Kouichiro~\textsc{Nakanishi}\altaffilmark{8,9}}

\author{Yiping~\textsc{Ao}\altaffilmark{3,17}}
\altaffiltext{17}{Purple Mountain Observatory and Key Laboratory for Radio Astronomy, Chinese Academy of Sciences, Nanjing 210034, China}

\author{Tohru~\textsc{Nagao}\altaffilmark{18}}
\altaffiltext{18}{Research Center for Space and Cosmic Evolution, Ehime University, 2-5 Bunkyo-cho, Matsuyama, Ehime 790-8577, Japan}

\author{Mariko~\textsc{Kubo}\altaffilmark{18}}

\author{Tsutomu~T.~\textsc{Takeuchi}\altaffilmark{19,20}}
\altaffiltext{19}{Division of Particle and Astrophysical Science, Nagoya University, Furo-cho, Chikusa-ku, Nagoya, 464-8602, Japan}
\altaffiltext{20}{The Research Center for Statistical Machine Learning, The Institute of Statistical Mathematics, 10-3 Midori-cho, Tachikawa, Tokyo 190-8562, Japan}

\author{Minju~\textsc{Lee}\altaffilmark{21,22}}
\altaffiltext{21}{Cosmic Dawn Center (DAWN), Jagtvej 128, DK-2200 Copenhagen N, Denmark}
\altaffiltext{22}{DTU-Space, Technical University of Denmark, Elektrovej 327, DK2800 Kgs. Lyngby, Denmark}

\author{Takuma~\textsc{Izumi}\altaffilmark{8,9}}

\author{Soh~\textsc{Ikarashi}\altaffilmark{23}}
\altaffiltext{23}{Junior College, Fukuoka Institute of Technology, 3-30-1 Wajiro-higashi, Higashi-ku, Fukuoka, 811-0295, Japan}

\author{Tohru~\textsc{Yamada}\altaffilmark{24,25}}
\altaffiltext{24}{Astronomical Institute, Tohoku University, 6-3 Aoba, Aramaki, Aoba-ku, Sendai, Miyagi 980-8578, Japan}
\altaffiltext{25}{Institute of Space and Astronautical Science, JAXA, 3-1-1 Yoshinodai, Sagamihara, Kanagawa Japan}



\maketitle

\begin{abstract}
    The ADF22 line survey reported detections of two high-$z$ line-emitting source candidates above 6-$\sigma$, both of which were shown to be spurious after follow-up observations. We investigate the detectability of far-infrared emitters in ALMA deep fields using mock observations by injecting artificial line-emitting sources into the visibility planes. We also discuss our investigation, conducted together with the ALMA operations team, of a possible technical problem in the original observations. Finally, we devise a method to estimate the \cplus\ luminosity function (LF) at $z \sim 6$, including a full analysis of signal contamination and sample completeness. 
    
    The comparison of pixel distributions between the real and mock datacubes does not show significant differences, confirming that the effect of non-Gaussian noise is negligible for the ADF22 datacube. Using 100 blank mock-mosaic datasets, we show 0.43 $\pm$ 0.67 false detections per datacube with the previous source-finding method. We argue that the underestimation of the contamination rate in the previous work is caused by the smaller number of datacubes, using only 4 real ADF22 datacubes. We compare the results of clump-finding between the time division mode (TDM) and frequency division mode (FDM) correlator datacubes and confirm that the velocity widths of the clumps in the TDM case are up to 3 times wider than in the FDM case. 

  Additional investigation into technical issues, specifically the `QA3 process', found no technical problems for both Cy-2 and Cy-4 data. Therefore, we confirm that false detections of high-SN ratios are unavoidable because clump-like structures already exist in large TDM datacubes. The LF estimation using our model shows that a correction for the number count is required, up to one order of magnitude, in the luminosity range of $\geq 5 \times 10^8$ \Lsun. Our reconstruction method for the line LF can be applied to future blind line surveys.
\end{abstract}

\section{Introduction}
    To reliably trace the evolution of the star-formation rate density (SFRD) as a function of cosmic time (e.g., \cite{madau2014}), we must be able to detect star-forming galaxies without biases such as dust obscuration or stellar mass, and unambiguously determine the spectroscopic redshifts for these galaxies. Historically, the Lyman-break method has been a powerful tool to study galaxies at high redshift, due to the relatively wide field and high sensitivity of optical observational facilities \citep{bouwens2012,bouwens2016,ono2018}. Dust-depleted star-forming galaxies are easily detected by the Lyman-break method. 
    Other methods, such as high-sensitivity spectroscopy or multiple observations, are very expensive in terms of observing time. However, high-sensitivity rest-frame far-infrared observations with the Atacama Large Millimeter/submillimeter Array (ALMA) now have the potential to move us into a new era through blind searches for (sub-)millimeter (mm) line emissions in traditional deep fields. 
    
\begin{figure*}
	\begin{center}
	\includegraphics[width=160mm]{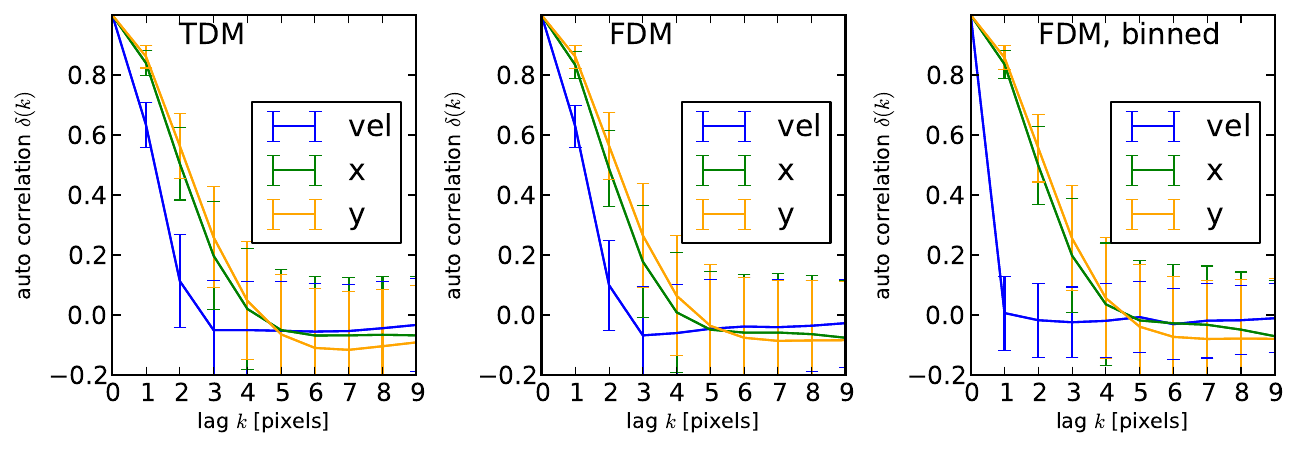}
  	\caption{The comparison of auto-correlation between TDM and FDM mode. The FDM observation shown in the right panel is binned from the frequency resolution of 1.129 MHz to 16.935 MHz so that the data have a comparable frequency, i.e., velocity resolution, to the TDM data. ALMA data exhibit auto-correlation on a scale of several frequency channels and angular resolutions, whereas the binned FDM data do not show correlation in the frequency/velocity domain (blue lines). Thus, ALMA data tend to have clumps on the scale of angular resolution and several channels, corresponding to a velocity width of $\sim$ 100 km s$^{-1}$. This implies that the datacube obtained with TDM has more clumps than the binned FDM datacube.}
	\label{fig:autocor}
	\end{center}
\end{figure*}

        We target the \cplus\ line emission, one of the main coolants of atomic gas (e.g.,\,\cite{wolfire1995}). Due to the high abundance of carbon, the large ionisation potential of singly ionised and atomic carbon (24.2 and 11.1\,eV), the low excitation temperature (92\,K), and the high critical density ($\approx 10^3$ cm$^{-3}$) of the transition, the line emission can arise from multi-phase gas (e.g.,\,\cite{hollenbach1989}). \cplus\ line emission is enhanced in star-forming regions, boosted by appropriate heating processes such as photoelectric heating of dust or cosmic-ray heating (e.g.,\,\cite{wolfire1995}). The redshift evolution of the \cplus\ 158-$\mu$m luminosity function (\cplus\ LF) thus probes the evolution of star-formation activity in the Universe.

    Blind line searches are essential to constrain the \cplus\ LF. A set of blind and serendipitous detections have already been used to constrain the \cplus\ LF (e.g.,\,\cite{matsuda2015, capak2015, aravena2016b, hayatsu2017}). Some of these \cplus\ emitters/candidates are faint at all other wavelengths. If real, the high-redshift star-forming activity must differ from that seen in the local Universe, lacking both rest-frame UV and dust continuum emission.

    In \citet{hayatsu2017}, we searched for millimeter line emitters in the 1.1-mm ALMA Deep Field in SSA22 (hereafter ADF22) survey data, which were taken in ALMA Cycle~2 by \citet{umehata2017}. In order to search faint emission-line sources, we use high-sensitivity data of 80 pointing fields, Field 2--Field 81 (see Fig.\,\ref{fig:mosaicID} in Appendix 2). Our line search, undertaken within a 5-arcmin$^2$ field, uncovered two CO line emitters associated with known galaxies at $z = 0.7$ and $3.1$, and two \cplus\ line candidates. These latter lines were both detected at $>$ 6$\sigma$ in spectral windows that contained no similarly significant negative features; both lines were also recovered in temporal subsets of the data, at lower significance.

    If \cplus\ is the correct line identification, then these galaxies dubbed ADF22-LineA and ADF22-LineB lie at $z = 6.0$ and $6.5$, respectively. They are faint in continuum at optical--infrared--mm wavelengths, with large equivalent widths. One negative line was detected at $>$ 6$\sigma$, albeit in a different and noisier spectral window, so Hayatsu \emph{et al.} predicted a likely contamination rate of $\approx 50$\%, meaning that we would expect one of the two candidates to be a real line emitter. The estimation of the contamination rate also considered the results of searches using different spectral smoothing widths. Nonetheless, the follow-up observations could not confirm the presence of line emitters at the same frequency \citep{hayatsu2019}. In this paper, we discuss the reasons for the false detections and how we used the data to set constraints on the \cplus\ LF.

    We examine the completeness and contamination for the ADF22 data observed in Cy.2 and discuss each possible explanation for the non-detection in the Cy.5 observation. We discuss technical and statistical issues in \S 2. In \S 3, we show the set of mock observational data and artificial sources. In \S 4, we present how to calculate completeness and contamination rates, and also a method for estimating the luminosity function by applying the outcomes. In \S 5, we show results and the detectability of FIR line emitters in ADF22 data. In \S 6, we discuss the case of abundant detection and show the result of the QA3 process. Finally, we present our conclusions in \S 7. Throughout the paper, we adopt the standard $\Lambda$-CDM cosmology with a matter density $\Omega_{\rm M} = 0.3$, a cosmological constant $\Omega_{\rm \Lambda} = 0.7$, and a Hubble constant $h = 0.7$ in units of $H_0 = 100$ km s$^{-1}$ Mpc$^{-1}$. The data are analyzed with the Common Astronomy Software Application (\texttt{casa}) ver.\,4.5.3 \citep{casa2022}.

\section{Investigations of the Cycle 2 Data and Possible Explanations of the False Detections}
    In the following, we discuss four reasons why the previous claims of detections of two line emitters at $z \sim 6$ may be spurious, despite the statistical significance being larger than 6$\sigma$. These possibilities are difficult to distinguish for ALMA users because the \texttt{casa} simulator is not fully utilized to reproduce real observations, i.e., the simulation for each effect is not supported\footnote{For instance, the system noise temperature $T_{\rm sys}$ parameter in the \texttt{casa} simulator is determined as an average value, but non-Gaussian noise can be caused by the difference in $T_{\rm sys}$ between the antennas.}. Therefore, we discuss the uncertainties related to the detectability of sources with a full analysis, including contamination and completeness, using mock observations and Monte Carlo simulations.

\subsection{Statistical Fluctuation After Convolution} 
    In the imaging process, visibility data in the UV-plane (two-dimensional spatial frequencies) will be Fourier transformed into a dirty map which is the brightness distribution convolved with the dirty beam or the point spread function (PSF). Therefore, pixels in a image data cube are not independent in the spatial and spectral domains (Fig.\ref{fig:autocor}), the number of independent pixels should be approximately two orders of magnitude less than the total number of pixels. If pixel values in a data cube follow the normal Gaussian distribution, the probability for an independent value exceeding $\pm 6\sigma$ will be $< 10^{-9}$ \footnote{946 pixels $\times$ 1896 pixels $\times$ 112 frequency channels $\times$ 4 spectral windows (SPWs) $\sim$ $10^9$.}. This means that fewer than 1 pixel can be above 6$\sigma$, i.e., the contamination rate for the original datacube is less than 50\%.

\begin{figure}
	\begin{center}
	\includegraphics[trim=100 0 100 0, width=80mm]{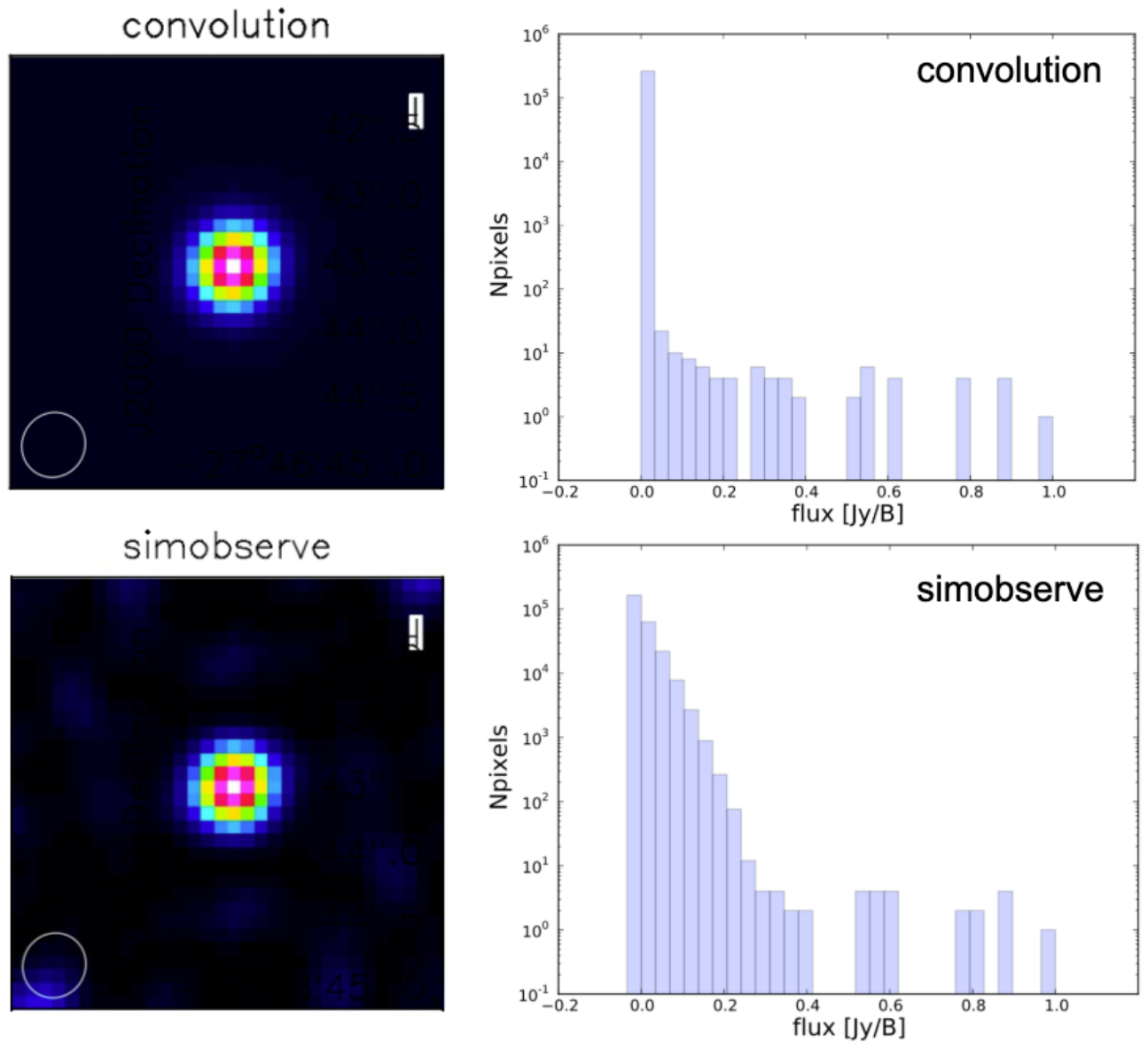}
    \caption{Appearance of an injected source. We inject a continuum emission as a point source with a peak flux of 1~Jy. The top panels show the result using \texttt{2dconvolve}, which performs a 2-dimensional convolution process at the source location. The bottom panels show the result using \texttt{cl.addcomponent} or \texttt{simobserve}, which process the visibility domain. These two methods exhibit different tendencies in the pixel distribution.}
	\label{fig:comp_source}
	\end{center}
\end{figure}

    However, \citet{gonzalez2017} showed that the occurrence of 6-$\sigma$ level pixels is not unusual due to statistical fluctuations in large ALMA survey datasets. Also, note that this process would affect the signal distribution (Fig.\ref{fig:comp_source}). We generate mock ADF22 datacubes considering the beam convolution in the UV-plane and the time division mode (TDM) setup of the correlator and discuss the probability of high-SN pixels arising in our case.

\subsection{Spectral Smoothing Process}
In ALMA observation, the 64-input correlator can be set either TDM or FDM and a large ALMA deep observation often requests the use of the TDM, which provides data with a wider velocity resolution of approximately a few tens of \kms, instead of the frequency-division mode (FDM). This setup reduces data size and is sufficient for both continuum and line searches. However, the TDM and the Hanning smoothing process in the spectral domain\footnote{Refer to the ALMA (Cycle 11) technical handbook \S5.6.2.} may lead to clumps with a galaxy-like velocity width of approximately $\sim$ 100\,\kms.

    In Fig.\,\ref{fig:autocor}, we present a comparison of the auto-correlation between TDM and FDM modes. The FDM observation is binned in frequency resolution from 1.129 MHz to 16.935 MHz to achieve a comparable frequency resolution to TDM data. The FDM data exhibits no correlation in the frequency domain due to the binning process, while the TDM data in the upper panel displays correlation in the frequency domain. Consequently, the number of independent voxels in the unbinned FDM data is 16 times larger than that in the TDM data. After the 16-channel binning (resulting in a ratio of 16.935 MHz to 1.129 MHz), the FDM data no longer exhibits spectral domain correlation. In this case, the number of channels becomes the same, and the number of independent elements for TDM data is approximately five times smaller than for the binned FDM data. Note that the correlated pixel extends to several pixels/channels (Fig.\,\ref{fig:autocor}).

    The effect on detectability can be confirmed by comparing the results of contamination checks using mock data generated by methods assuming TDM/FDM correlator.

    The spectral-smoothing process in the source-extraction procedure may also enhance the clump-like structure in the spectral domain. The best smoothing width would be the same as the FWHM of the targeting line (Fig.\ref{fig:smoothing} in Appendix 1 for the calculation). We then show the comparison of the contamination rate by separating the results of different smoothing widths. Note that such a smoothing process should affect both the positive and negative directions, though it may also enhance the non-Gaussian effect, which arises only in the spatial domain.

\begin{figure}
	\begin{center}
	\includegraphics[width=80mm]{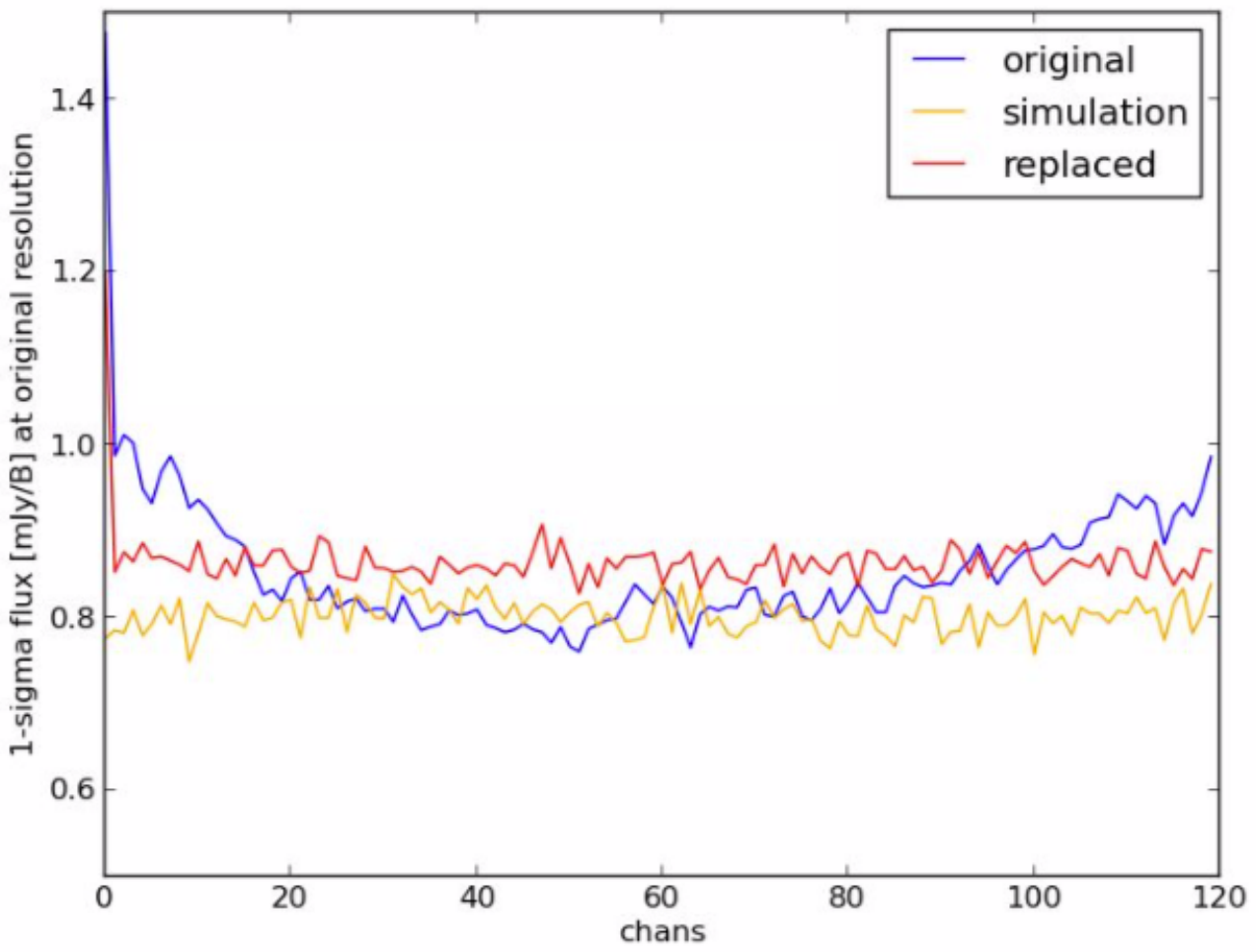}
  	\caption{Comparison of the RMS noise for simulated and original data as a function of frequency channel. The blue and red lines show different generation processes; original and replaced. All mock data sets successfully reproduce the RMS noise level.}
	\label{fig:comp_obs_rep_sim}
	\end{center}
\end{figure}

\subsection{Non-Gaussian Noise} 
    Non-Gaussian noise is caused by, e.g., amplitude error for antennas or convolution error, and is recognized as unnatural patterns or ripples in the image plane. The effect depends on the comprehensive observational conditions, which are difficult to parameterize. The problem is that we assume the Gaussian distribution of the noise in the interferometric data and look for a sign of the presence of a source as a positive excess from the Gaussian distribution. We discuss the effect of the noise by comparing the pixel distributions and the results of contamination checks between the original (containing non-Gaussian noise) and mock mosaic data built with pure Gaussian noise. Also, visual inspection is important to recognize the non-Gaussian effect. We show the image of the original/smoothed data and re-check if such a tendency exists around the detected candidates.

\subsection{Technical Problems} 
    We investigated whether a technical problem occurred during the Cycle 2 observations. We have examined all the observing logs and the data scan by scan and re-run the pipeline with a more updated (and reliable) version. We have further checked whether a wrong coordinate could have been assigned in one or more of the fields contained in the mosaic image. We have checked if the phase tracking center measured for each field was wrongly assigned to the next field while the antennas pointed in the correct direction.  
    We checked the equatorial offsets for one of the observations as a function of time and each other using an executable and found no missing directions using the function {\sc aU.plotDelayUpdates}. Additionally, the conclusion of the quality assurance process, QA3, carried out by the ALMA observatory staff, stated that no technical issues were found.  

\begin{figure}
	\begin{center}
	\includegraphics[trim=70 180 70 150, width=80mm]{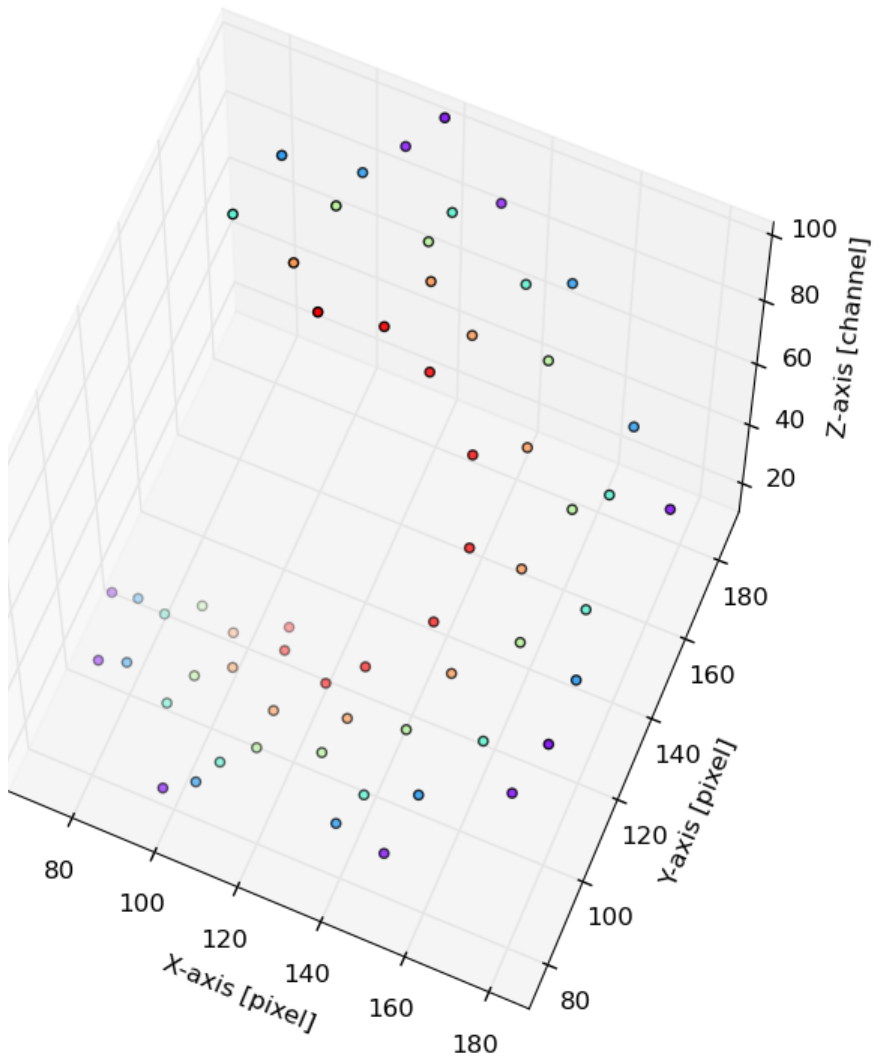}
    \caption{An example of input positions of the artificial sources. The color indicates different primary beam (pb) values. The pb value increases from blue to red. We select the input so that there is no overlap for each of the sources.}
	\label{fig:injectpoint}
	\end{center}
\end{figure}

\section{The Mock Observations}

\begin{figure}
    \begin{center}
	\includegraphics[trim=0 0 0 0, width=80mm]{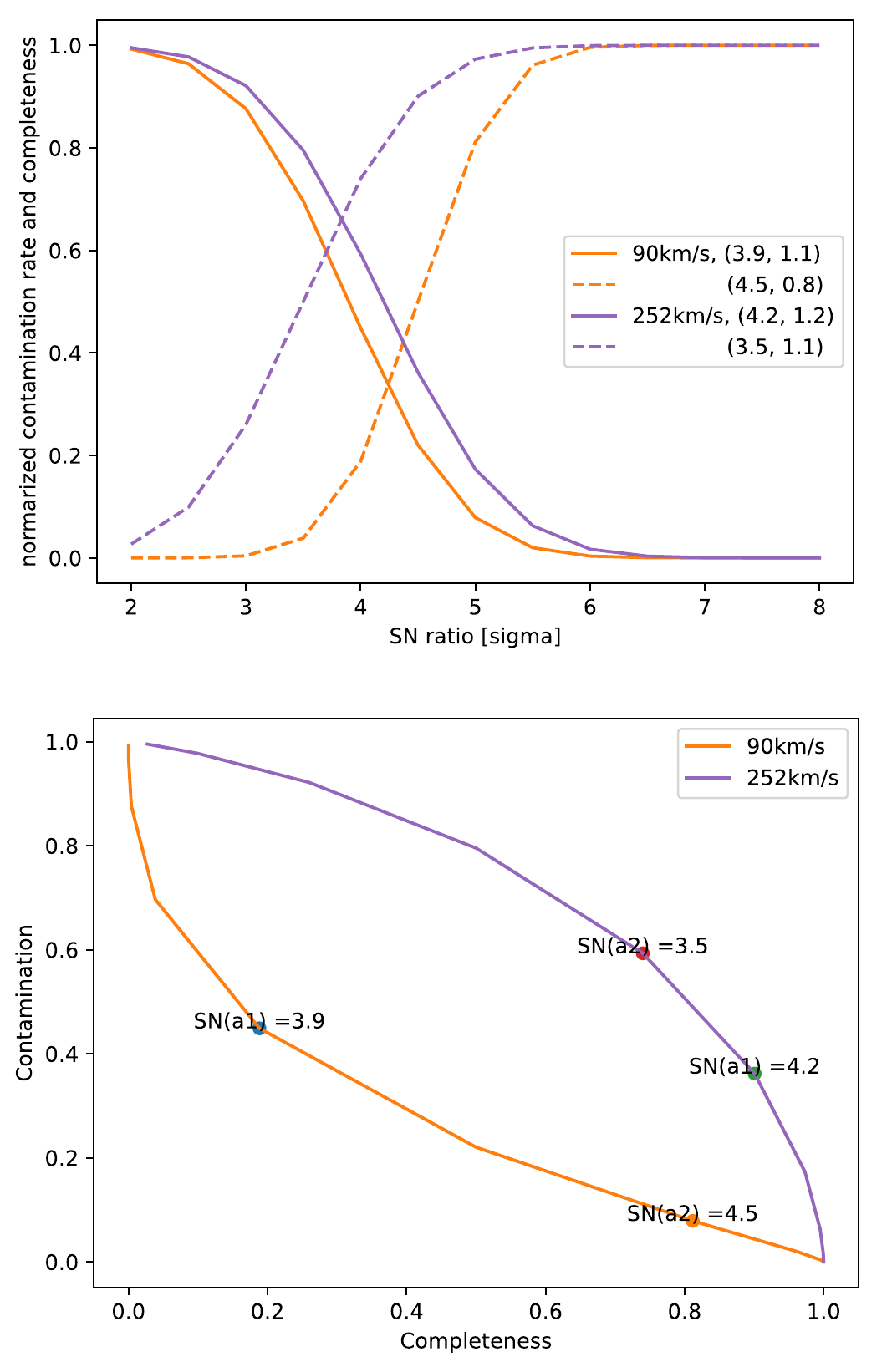}
    \caption{(Top) An example of the contamination rate (solid line) and completeness (dashed line) as a function of peak SN ratio. The contamination rate is normalized and cumulative. The orange and purple lines represent small and large velocity widths (90\,\kms\ and 252\,\kms\, respectively) with corresponding smoothing values (5 channels smoothing and 14 channels smoothing). The numbers in parentheses show the parameters \(a\) and \(b\) used in Eqs.\,\ref{eq:Qfit} and \ref{eq:Cfit}. For the 90\,\kms\ case, \(a_1 < a_2\), while for the 252\,\kms\ case, it's vice versa, meaning the larger velocity width achieves higher completeness.

    (Bottom) When the functions are combined using the SN ratio as a parameter, we see the detectability of the source is classified into three categories based on the inflection points. The threshold peak SN values are 3.9\(\sigma\) and 4.5\(\sigma\) for the 90\,\kms\ case and 3.5\(\sigma\) and 4.2\(\sigma\) for the 252\,\kms\ case, respectively.}
    \label{fig:exampleCandQ}
    \end{center}
\end{figure}

    We consider two ways to generate mock ALMA data:
\begin{description}
    \item \textbf{(1)} In order to generate data that have the same properties as the template data but with pure Gaussian noise, we replace the noise distribution with Gaussian in the UV-plane of the original data. These latter data can be used for the computation of the contamination rate. We compare the SN distribution between the replaced data and the original data and then discuss the effect of non-Gaussian noise.
    
    \item \textbf{(2)} By using \texttt{simobserve} in \texttt{casa}, we compare the difference between Time Division Mode (TDM) and Frequency Division Mode (FDM) observation. In this case, the dataset does not necessarily reproduce the properties of real observations. We generate two sets of single-field data which have exactly the same setup but with either FDM or TDM observation.
\end{description}

    In Fig.\,\ref{fig:comp_obs_rep_sim}, we show the comparison between the two ways using a test datacube of a single field. We describe how to generate these mock data in the following subsection. 

\begin{figure}
	\begin{center}
	\includegraphics[trim=0 0 0 0, width=80mm]{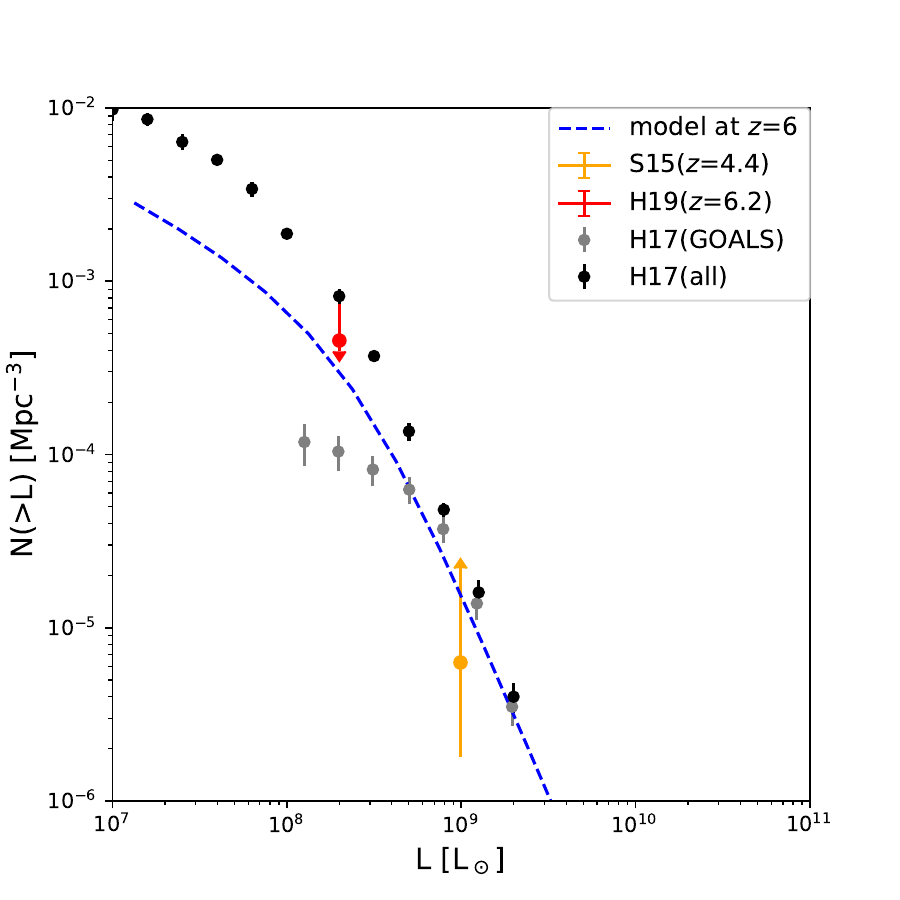}
  	\caption{We plot the observed \cplus\ LFs and an example of our model. The blue dotted line shows the model of \cplus\ LF at $z = 6$: it has a double-power law shape and redshift evolution of the amplitude from the local LF \citep{hemmati2017}. The orange and red arrows are observational lower and upper limits from \citet{swinbank2012} and \citet{hayatsu2019}, respectively. Black and grey dots are constraints at the local Universe from \citet{hemmati2017}, with and without considering the sample completeness and the effective volume of the survey, respectively.}
	\label{fig:modelLF}
	\end{center}
\end{figure}

\subsection{Replaced Data}
    We measure the RMS value and the mean value of the noise in the visibility plane and replace the values with pure Gaussian noise. Then we perform the Hanning smoothing process to reproduce the correlation between the frequency channels.
    We replace the tables of both the real and imaginary parts using a \texttt{casa} command \textsc{tb.putcol} and multiply the RMS values by $\sqrt{3}$ since the smoothing process reduces the values by a factor of $\sqrt{3}$. We perform this process for each field, and then concatenate the elemental fields to reproduce the mosaic field. The generated visibility data are converted into the 'dirty' image cube using \textsc{tclean}. We generate 25 mock mosaic data for 4 SPWs to check the effect of statistical fluctuation. Additionally, we prepare 320 single-pointing data to examine the non-Gaussian effect. The smoothing and normalizing processes are performed as for the original data. Note that the replacement process can reproduce the flagging information without further treatment.
    
    Note that we do not expect faint sources in the data to cause non-uniformity, as the pixel distribution of the original data closely aligns with a Gaussian profile (Fig.\,20 in Appendix 4).

\begin{figure}
	\begin{center}
	\includegraphics[trim=0 0 0 0, width=85mm]{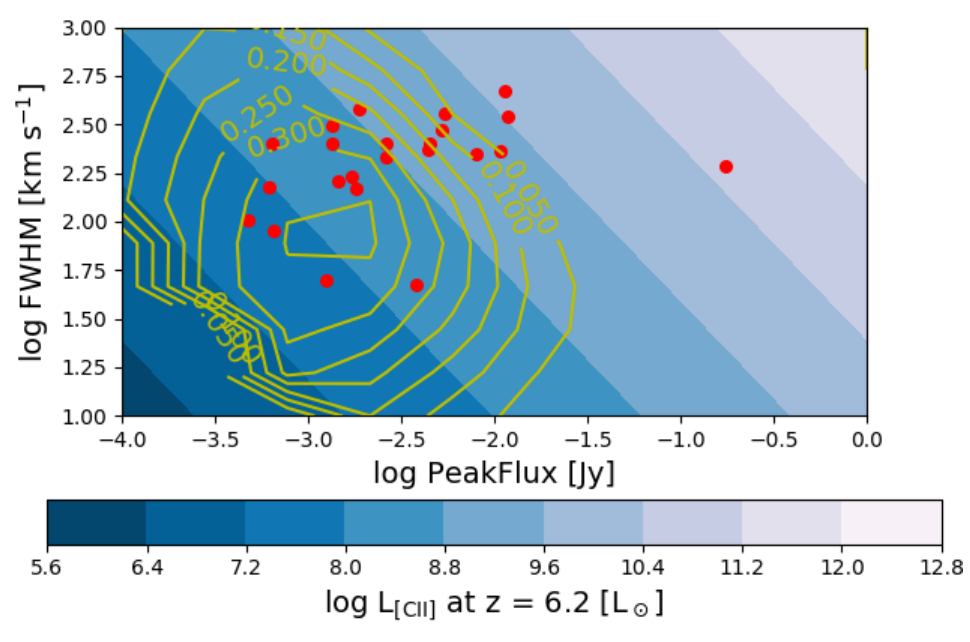}
    \caption{The distribution of FWHM and peak flux. Red points and the yellow contour show known \cplus\ emitters at $z = 4.4$--$7.1$ \citep{kohandel2019} and our model, respectively. The observational result shows a correlation, though the number of samples is small. Therefore, we assume the distribution and obtain the weighting function.}
	\label{fig:modeldist}
	\end{center}
\end{figure}

\subsection{Simulated Data}
    In order to check how the detectability is affected by the choice of correlator modes, we generate mock single-pointing data for both TDM and FDM modes using {\sc simobserve}, with pure Gaussian noise, with Hanning smoothing applied, as in the real observations. For simplicity, we choose the same observing time of 4.5 min and noise value for each field, because observing time affects the UV-plane coverage primarily. The velocity resolution of the cube was set to be 36\, km\,s$^{-1}$. Note that the UV-plane coverage does \textit{not} need to be reproduced for this test\footnote{To reproduce UV-plane coverage, one needs to match the observed time and hour angle for each scan, e.g., 720 sets of observations for this case, and concatenate the elemental observation.}. We use each field as the test, i.e., 320 data sets for TDM and FDM, respectively. We run {\sc simobserve} for each 9 scans by preparing the information of pointing, configuration, observing time, and starting time.  

\begin{table*}
    \caption{The normalized weighting values discussed in \S4.2.}
    \centering
    \label{tab:weighting}
    \begin{threeparttable}
    \begin{tabular}{c|ccccccccccc}\hline
    \multicolumn{1}{c|}{FWHM} &  \multicolumn{11}{c}{peak SN ratio [$\sigma$]}\\
    \multicolumn{1}{c|}{[chan]}&2.0&2.5&3.0&3.5&4.0&4.5&5.0&5.5&6.0&6.5&7.0\\ \hline \hline
2&1.0&0.55&0.36&0.26&0.2&0.16&0.12&0.1&0.086&0.073&0.06\\
4&0.76&0.43&0.28&0.2&0.15&0.12&0.095&0.076&0.063&0.053&0.043\\
6&0.61&0.35&0.23&0.16&0.12&0.091&0.072&0.059&0.047&0.038&0.032\\
8&0.5&0.28&0.19&0.13&0.096&0.075&0.055&0.044&0.036&0.028&0.023\\
10&0.41&0.24&0.16&0.11&0.078&0.058&0.045&0.035&0.027&0.022&0.016\\
12&0.35&0.2&0.13&0.09&0.064&0.049&0.036&0.028&0.02&0.017&0.013\\
14&0.3&0.17&0.11&0.077&0.055&0.04&0.028&0.022&0.017&0.013&0.0095\\
16&0.27&0.15&0.098&0.066&0.046&0.032&0.023&0.018&0.013&0.011&0.0075\\
18&0.23&0.14&0.086&0.057&0.039&0.028&0.018&0.014&0.01&0.0073&0.0057\\
20&0.2&0.12&0.073&0.048&0.032&0.022&0.017&0.012&0.0081&0.0063&0.0046\\
22&0.18&0.11&0.064&0.042&0.028&0.018&0.014&0.009&0.0062&0.0049&0.0036\\ \hline
    \end{tabular}
    \begin{tablenotes}
    \item[ ] The channel width corresponds to $\sim$ 18 \kms.  
    \end{tablenotes}
    \end{threeparttable}
\end{table*}

\subsection{Artificial Sources}
    We examine at least three ways to inject artificial line emitting sources in the ALMA datacube. A simple way is to generate convolved line emitters and add them to the image datacube. This can be processed using, e.g., \texttt{ia.convolve2d} for each channel. To inject sources in the UV-plane, one can use \texttt{simobserve} by specifying the peak brightness with a parameter \texttt{inbright}. In this case, injection to real/replaced datacube is impossible. We thus inject artificial sources to the UV-plane using \texttt{cl.addcomponent}. We define the source locations and calculate the flux value for each data slice. We prepare a source model with \texttt{cl.addcomponent} and Fourier-transform the model using \texttt{ft}. Since the information of the model is stored in a column named {\sc model\_data}, we need to inject the model to the {\sc corrected\_data} column. Then we use \texttt{uvsub} with a parameter {\sc reverse=True} since the command is usually to subtract the model in UV-plane.  
    Fig.~\ref{fig:comp_source} compares the results of the two ways. Note that the results using \texttt{simobserve} and \texttt{cl.addcomponent} are identical. We have decided to inject sources in the visibility plane to make it as realistic as possible.

\subsection{Source Injection}
    In this subsection, we define the parameter range of the artificial sources and how to define the input position. The minimum FWHM of the artificial sources is set to the same as the velocity resolution $\sim$ 36\,\kms. The maximum FWHM is set to be 400\,\kms, the same as the smoothing parameter. We set the minimum SN ratio of the injection source to be 2$\sigma$ because the peak SN ratio at the original spectral resolution is $\sim$ 2$\sigma$ \citep{hayatsu2017}. The maximum input SN ratio is set to be 7$\sigma$ to discuss the detectability of $>$ 6$\sigma$ emitters.  
    This parameter range covers the luminosity range of $10^8$--$10^9$ \Lsun. The calculation to convert flux to the \cplus\ luminosity, $L_{\rm [CII]}$, using luminosity distance $D_{\rm L}$, observed frequency $\nu_{\rm o}$, velocity-integrated flux $S^{\rm v}$ is written as (e.g.,\,\cite{carilli2013})

\begin{equation}
    \frac{L_{\rm [CII]}}{L_\odot} = 1.04 \times 10^{-3} \left( \frac{D_{\rm L}}{\rm Mpc} \right)^2 
    \frac{\nu_{\rm o}}{\rm GHz} \frac{S^{\rm v}}{\rm Jy~km~s^{-1}}. 
    \label{eq.flux_Lline}
\end{equation}    
    
    Note that very high luminosity emitters are not likely identified even with 100\,\% completeness because of the extremely small number density. Using the observational upper and lower limits from \citet{hayatsu2019} and \citet{swinbank2012}, we roughly estimate $\sim$ 10$^{-2}$--1 ($\sim$ 10$^{-3}$--0.1) \cplus\ emitters with $\sim 10^{8}$--$10^{9}$ \Lsun\ can be detected within the ADF22 survey volume of 2200 Mpc$^{-3}$.    
    The bin size of the peak SN is set to be 0.5, which is the same as the inclination level to the search using {\sc clumpfind} \citep{williams1994}. Thus the input S/N ratio varies 2.0, 2.5, ... 7.0$\sigma$ (11 types). The FWHM varies as 36, 72, ..., 396 \kms (11 types). Therefore, we inject artificial sources with 121 parameter sets of the artificial source. We then place 10 sources each at the position of pb = 0.81, 0.84, ..., 0.99 and channels = 9, 19, 29, ... 119. Then 84 positions in total. We select the positions such that there is no overlap. 

    In Fig.\,\ref{fig:injectpoint}, we show an example of the injection points for a single field.  
    To evaluate the statistical error, we repeat the same process 100 times with different fields. We inject 1,016,400 sources in total.  
    The parameter set is converted to intrinsic peak flux and FWHM by considering the injected pb-values. The flux density can be linearly converted to the intrinsic luminosity by using Eq.\,\ref{eq.flux_Lline}, assuming the average \cplus\ redshift to be $z = 6.2$. 

\section{The Detection Methods}
\subsection{Detection Classes}
    After finding clumps, we need to define the detection threshold and evaluate its reliability. For example, detection with a higher S/N ratio would indicate higher completeness and a lower contamination rate, suggesting more reliable detection. Contamination rate is calculated by comparing the number of pixels/clumps above a threshold value between the source-containing data and blank-field data. In \citet{hayatsu2017}, the number of clumps in the blank-field data is calculated using inverted data; the datacube is multiplied by -1.0. In this case, we evaluate the contamination rate with four datacubes. To increase the number of samples, we update the calculation by using `blank' mock data, i.e., the mock datacube with only noise. Here we define the contamination rate $Q$ for a certain parameter $x$, e.g., flux, line width, or size, as 
\begin{equation}
    Q (x) = N_{\rm mock} (x) / N_{\rm real} (x), 
\end{equation}
    where $N_{\rm real} (x)$ and $N_{\rm mock} (x)$ are the cumulative number of clumps in the real and the blank mock field without injected sources, respectively, as a function of $x$. Providing an example would be illustrative here. Let us set $x$ to be SN ratio and suppose that we are interested in the contamination rate at SN = 4$\sigma$. If the number of clumps above SN = 4$\sigma$ in the blank mock data is $N_{\rm mock}$ = 50 and that in the real data is $N_{\rm real}$ = 100, respectively, then $Q(x)$ is 0.5, meaning that half of the 100 clumps in the real data are fake (noise). We consider the region where $N_{\rm mock} \leq N_{\rm real}$ and $N_{\rm real} > 0$\footnote{Note that fidelity $P(x)$ in \citet{aravena2016} is defined by $P(x) = 1 - Q(x)$.}.
    We choose $x$ as a parameter set of (peak SN, FWHM). Since we assume high-$z$ FIR line emitters are detected as point sources in ALMA deep survey data, the parameter set is directly connected to the line luminosity (see Eq.\ref{eq.flux_Lline}). $Q(x)$ is estimated with smoothed SN cubes, and the smoothing parameter corresponds to the FWHM as expressed in \S appendix 1. 

\begin{figure}
	\begin{center}
	\includegraphics[trim=0 0 0 0, width=80mm]{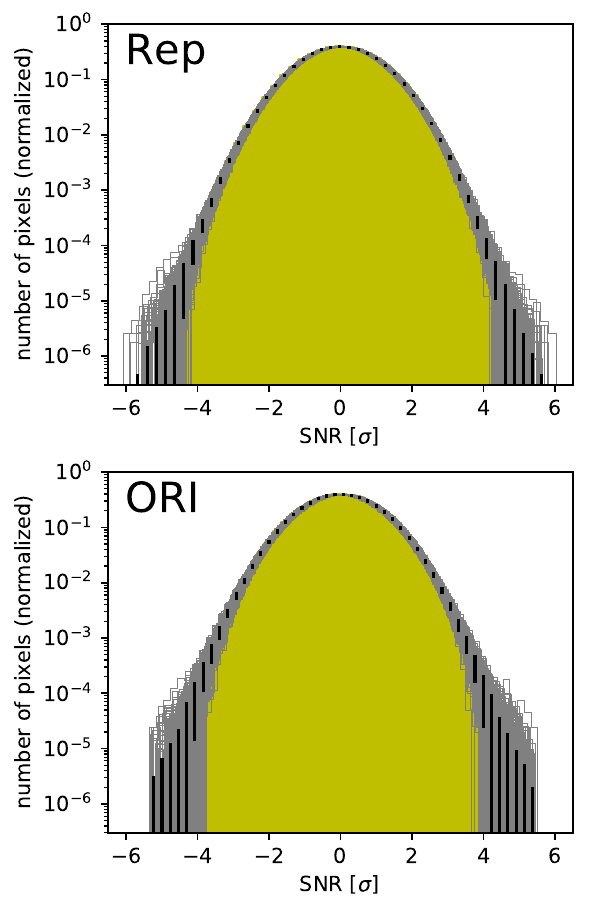}
  	\caption{Pixel distributions for single field data for mock data and real data. The upper panel shows the mock data generated by the replacing process. The bottom panel shows the original data. To avoid the positive excess caused by real sources, we mask the positions of the known continuum and emission sources in the size of a beam for both data. We also smoothed the datacube in the spectral axes with a width of 12 to enhance the SN ratio of the clumps. Grey lines show the individual histograms, the yellow shaded region shows the mean shape, and the black lines show 1$~\sigma$ errors. If the non-Gaussian effects from real observation are not negligible, we expect to see a difference between the two, whereas there is no statistical difference between them.}
  	\label{fig:pixeldist_res2}
	\end{center}
\end{figure}

    We calculate the value of completeness $C(x)$ by injecting an artificial source and checking if the source is detected by the source-extracting process. We do this by using Monte-Carlo simulations with artificial sources. $C(x)$ is defined by the ratio of the number of detected artificial sources $N_{\rm out}$ to the total number of injected sources $N_{\rm inj}$ at the certain property $x$ of the source. Then we define $C(x)$ as  
\begin{equation}
    C (x) = N_{\rm out} (x) / N_{\rm inj} (x). 
\end{equation}
    For example, if we inject 100 sources of SN = 4$\sigma$ and detect 20 clumps of them with our method, $C({\rm SN} = 4 \sigma)$ = 0.2. In this case, we miss 80\% of 4$\sigma$ sources with the method.  

\begin{figure}
	\begin{center}
	\includegraphics[trim=0 0 0 0, width=80mm]{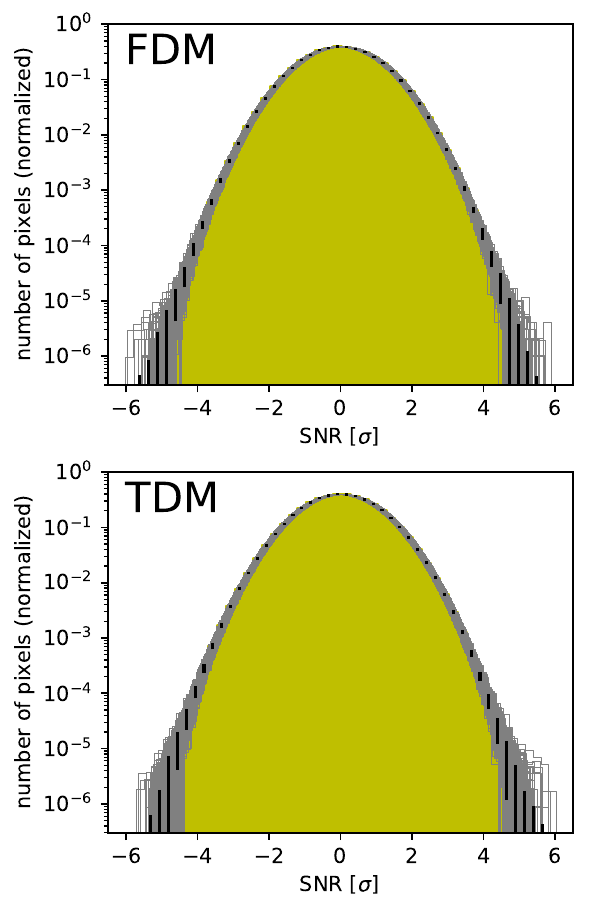}
  	\caption{Pixel distributions for mock FDM and TDM data. Upper and lower panels show the FDM and TDM observation, respectively. These mock data sets are generated using {\sc simobserve}. We also smoothed the datacubes along the spectral axes with a width of 12 to enhance the SN ratio of the clumps. Grey lines show the individual histograms, the yellow shaded region shows the mean shape, and the black lines show 1$~\sigma$ errors. Between TDM and FDM data, no statistical difference is found in the pixel distribution.}
  	\label{fig:pixeldist_res}
	\end{center}
\end{figure}

    After the injection of the source, we trim a region around the artificial source and smooth it spectrally. We then normalize the frequency data slice with the RMS values and execute {\sc clumpfind} \citep{williams1994}. The RMS values are measured in advance using the blank field data.  
    
    The results of $Q(x)$ and $C(x)$ can be approximated with the error function (e.g., \cite{hatsukade2016}):

\begin{eqnarray}
    Q(x) &\sim & (1 + {\rm erf}( \frac{x - a_{1}}{b_{1}})) / 2, \label{eq:Qfit}\\
    C(x) &\sim & (1 - {\rm erf}( \frac{x - a_{2}}{b_{2}})) / 2, \label{eq:Cfit}
\end{eqnarray}
    where $a_{1}$, $a_{2}$, $b_{1}$, $b_{2}$ are fitting parameters and $x$ is the property of the source. These decreasing/increasing functions have an inflection point ($a_{1}$ and $a_{2}$), and thus the reliability of the detection becomes robust at a certain threshold of S/N ratio (Fig.\,\ref{fig:exampleCandQ}). With the inflection points, we can discuss the reliability of the detection for each source property. For example, when we combine $Q(x)$ and $C(x)$ as shown in Fig.\,\ref{fig:exampleCandQ} lower panel, the function has one or two inflection points. The points represent boundaries of the reliability of the detection. If $a_{1} < a_{2}$, the reliability is classified as follows:

\begin{figure}
	\begin{center}
	\includegraphics[trim=0 0 0 0, width=80mm]{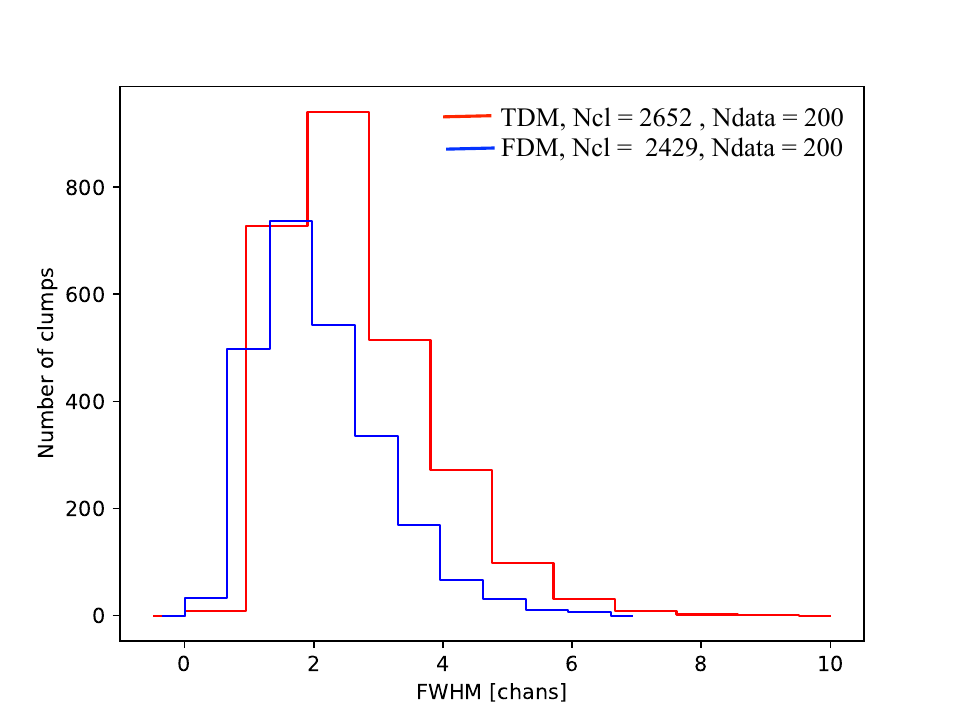}
  	\caption{The comparison of the detected number of {\sc clumpfind} \citep{williams1994} for TDM and FDM single field as a function of FWHM. The peaks are 1.65 and 2.38 for the FDM and TDM case, respectively. The TDM case tends to have clumps with a larger velocity width.}
  	\label{fig:hist_clumpTDMFDM}
	\end{center}
\end{figure}

\begin{description}
    \item {\bf Category I.} $a_{2}$ $<$ SN; ~the clump can be treated as a candidate and can be used for calculating the LF.
    
    \item {\bf Category II.} $a_{1}$ $<$ SN $\leq$ $a_{2}$; ~the clump cannot be treated as a candidate but can be used for LF estimation.    
    \item {\bf Category III.} SN $\leq$ $a_{1}$; ~the clump cannot be treated as a candidate nor can it be used for LF estimation due to incompleteness and high contamination rate. 
\end{description}

    Although a blindly detected source contaminated with noise cannot be treated as a real source, we can use the result of such detection to estimate or constrain the LF after the appropriate correction of the contamination and completeness contributions. 
    
\subsection{The LF Reconstruction}

\begin{figure}
	\begin{center}
	\includegraphics[trim=0 0 0 0, width=80mm]{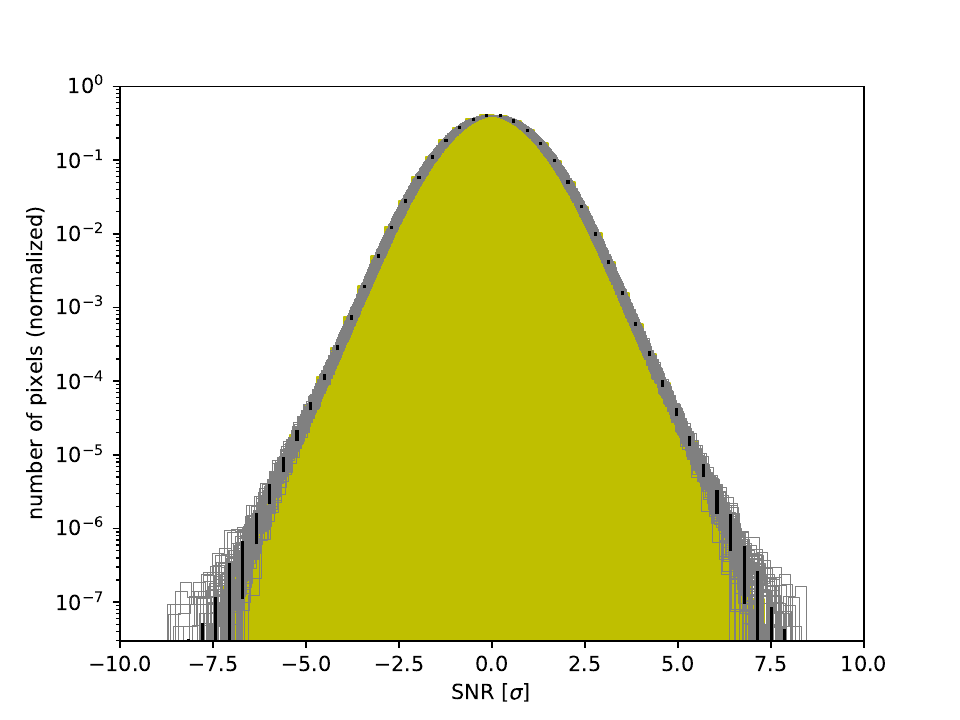}
  	\caption{Pixel distributions for mock mosaic data sets. We smoothed the datacube in the spectral domain with a width of 12 to enhance the SN ratio of the clumps. Grey lines show the individual histograms, the yellow shaded region shows the mean shape, and the black lines show 1$~\sigma$ errors. 6$~\sigma$ level pixels in the blank field are not unusual for ALMA mosaic data.}
	\label{fig:pixeldist_res3}
	\end{center}
\end{figure}

    The corrected detection number $N_{\rm cor}$ for each parameter set (FWHM, peak flux) is calculated using the detected number in the real datacube, $N_{\rm det}$, contamination rate $Q$, and completeness $C$:
    \begin{equation}
    N_{\rm cor} = N_{\rm det} (1 - Q)/C. 
    \end{equation}
    Note that $N_{\rm det}$ is not $N_{\rm real}$ because the latter is defined as a cumulative function. For example, when $N_{\rm det}$, $N_{\rm real}$, $N_{\rm mock}$, $N_{\rm inj}$, and $N_{\rm out}$ = 50, 120, 60, 100, and 60, respectively, then $N_{\rm cor}$, the actual total number of sources in the field, is 41.7. In this case, half of the original 50 clumps are fake and we miss 40\% of sources.
    
    Ideally, when the number of detections is large, and the interested parameter space has a sufficiently low contamination rate and high completeness, the source properties should trace the real distribution. However, the detected source property should be biased by the detection limit and the method. Therefore, we `weight' the $N_{\rm cor}$ values for each parameter set in each luminosity bin by assuming the distribution of the source property.  
    
    Also, we calculate a survey volume for each parameter set by considering the difference in frequency range used for the search and obtain the effective survey volume for each luminosity bin by `weighting'. Note that the survey volume of the unsmoothed datacube is 2200 Mpc$^{-3}$ and the typical number of channels of the datacube is 120. For example, if the smoothing width is 8 channels, i.e., 144 \kms, the survey volume for the sources with FWHM $\sim$ 144 \kms\ is 2053 Mpc$^{-3}$.
    
    Therefore, the corrected number density for each luminosity bin can be derived by the weighted $N_{\rm cor}$ and the weighted effective survey volume. For example, for parameter sets (FWHM, peak flux) = (216 \kms, 2 mJy) and (108 \kms, 4 mJy), having the same luminosity, let's say $N_{\rm cor}$ values are 3 and 2, and the weighting values are 0.2 and 0.1, respectively. In this case, the corrected number density at the luminosity is 2.67/2016.3 = 1.3 $\times$ 10$^{-3}$ Mpc$^{-3}$.
  
    In order to obtain the distribution in the FWHM-peak flux plane, we assume the \cplus\ LF at $z \sim 6$. After obtaining the LF, we calculate the FWHM value corresponding to each luminosity value assuming the observed correlation between \cplus\ luminosity and FWHM. The latter implies the Tully-Fisher relation \citep{kohandel2019}. We then calculate the peak flux value with the obtained FWHM and luminosity.

    Fig.\,\ref{fig:modelLF} shows the cumulative \cplus\ LF in the local Universe and at high-redshift. The dashed lines show the model example and the points show the observations \citep{swinbank2012, hemmati2017, hayatsu2019}. The estimated \cplus\ LFs at $z$ = $0.0023$--$0.076$ \citep{hemmati2017} are derived from follow-up observations of samples from the Great Observatories All-sky LIRG Survey \citep{diaz2013}. \citet{hemmati2017} estimate with a correction to the effective volume and sample number from a completeness analysis and the result is 2--3 dexes higher than the values at $z$ $\leq$ 0.05 \citep{swinbank2012}. 
    We correct the values from the differential to cumulative by multiplying the luminosity bin (dlogL$_{\rm [CII]} =$ 0.2 dex) and by adding the values to the lower luminosity bin. The delivered LF is better fitted with a double power-law than with a Schechter function. The double power-law luminosity function is 

\begin{figure}
	\begin{center}
	\includegraphics[trim=10 10 10 10, width=80mm]{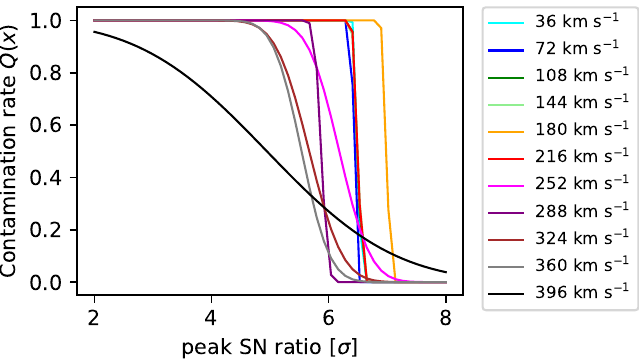}
  	\caption{The result of the calculation of contamination using real and mock mosaic data as a function of peak S/N ratio. The different colours show the different smoothing parameters, meaning different velocity widths of the detected clumps. The inflexion points of the functions have high values (4.9--6.5$\sigma$), meaning most of the sources in the ADF22 field are contaminated with noise.}
	\label{fig:contamination}
	\end{center}
\end{figure}

    \begin{equation} 
    \phi(L) = \phi_* \left((L/L_*)^\alpha + (L/L_*)^\beta \right)^{-1},
    \end{equation}
    where $\alpha = 2.36\pm 0.25$, $\beta = 0.42\pm 0.09$, $L_* = (2.173\pm 0.743)\times 10^8 {\rm L_\odot}$, $\phi_* = (0.003\pm 0.002)~{\rm Mpc^{-3}}~({\rm dlog L_{[CII]}})^{-1}$.  

    As for the constraint at high-$z$, we indicate the lower limit at $z$ = $4.4$ based on two serendipitous detections in the ALESS survey \citep{swinbank2012} (hereafter S12) and the upper limit from \citep{hayatsu2019} (hereafter H19). We do not consider the estimation from \citet{capak2015} at $z$ = $5$--$6$. The estimation is derived from follow-up observations and correcting the volume with the number density of Lyman break galaxies. 

    We also do not indicate the estimation provided by the result of ASPECS \citep{aravena2016b}, which is based on the assumption that all \cplus\ candidates which have a contamination rate of 60\% are real \cplus\ emitters at $z$ = 6--8. This constraint can be regarded as an upper limit. 

    Owing to the limitation of observational constraints at high-$z$, we make the simplest assumption and define the shape of the high-$z$ LF as a double power-law with the same parameters except for the amplitude. The amplitude (hereafter $\phi_*$) is determined as a result of the blind search. The redshift evolution of the amplitude is defined by 
\begin{equation}
\phi_*(z) = \phi_{*0} \cdot (1 + z)^\gamma
\end{equation}
and $\gamma < -0.3$ satisfies the constraint from S12 and H19, naively explaining that the evolution of the \cplus\ LF is determined by the merging process (change of the number) rather than the star-formation process (change of the luminosity).
In Fig.~\ref{fig:modeldist}, we show the distribution of FWHM and peak flux for high-$z$ \cplus\ emitters from our model and observation. 
In Tab.~\ref{tab:weighting}, we show the values of the weighting. 

Using the assumed distribution, we obtain the luminosity distribution at high-$z$, but the amplitude (number density) is unknown. We attempt to determine the amplitude from the result of our line survey.


\section{Results}
\subsection{Explanation of the False Detection}

\begin{figure}
	\begin{center}
	\includegraphics[trim=0 0 0 0, width=80mm]{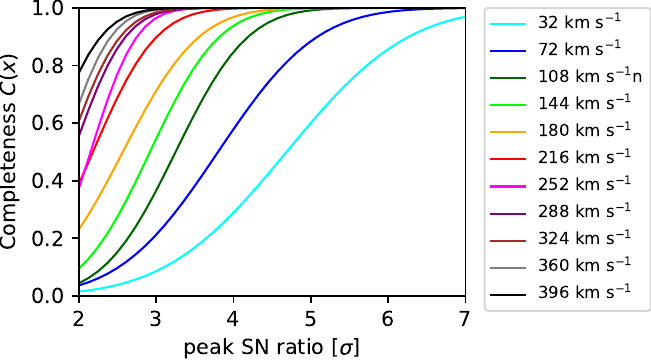}
  	\caption{The result of fitting of completeness analysis as a function of S/N ratio. The smaller velocity width has the larger inflection value, indicating incompleteness of clumps with small velocity width.}
	\label{fig:completeness}
	\end{center}
\end{figure}

\subsubsection{Non-Gaussian effect}
In order to investigate the noise properties in detail, we compare the pixel distribution between the real and mock data (Fig.\,\ref{fig:pixeldist_res2}). The mock data is generated by the replacement process, thus the mock data does not contain non-Gaussian noises. To distinguish the non-Gaussian effect from statistical fluctuations, we use a total of 320 single field data sets. The positions of the known sources in the real data are masked to avoid confusion by real sources. To enhance the SN ratio of the clump, we also smooth the datacube in the frequency domain with a width of 12 channels. The maximum and minimum SN values extend $\pm$ 4--6$~\sigma$ for both real and mock data, and there is no statistical difference between them. Therefore, we conclude that the non-Gaussian noise does not affect the false detection.  

We also find overdensities of the detected clumps in some of the replaced data. We assume that this noise tendency does not significantly affect the calculation of the contamination rate because the S/N distribution and clump number counts do not differ greatly from the original data. However, since we discovered this overdensity in the replaced data after the QA3 process, one should remain cautious of any potential instrumental effects that cause this noise tendency. 

\subsubsection{TDM and FDM}
By using {\sc simobserve}, we generate mock TDM and FDM observation data and compare their noise properties. The FDM data is binned in frequency axes to match frequency resolution to the TDM data. Therefore, there is no correlation in the frequency domain in the FDM data. We smooth the data in spectral axes with 12 channels.

We see no statistical difference in the pixel distributions (Fig.\,\ref{fig:pixeldist_res}), however, there is a difference in the distributions of clumps as a function of FWHM (Fig.\,\ref{fig:hist_clumpTDMFDM}). The peaks of the distributions are 1.65 and 2.38 for the FDM and TDM case, respectively. We also find a positive excess for the TDM case. The tendencies express that the noise clumps in the TDM data have larger velocity widths compared to those in the FDM data.

For an example of one realization, we detected ten clumps in each case and confirmed that the peak S/N ratios of the detected clumps were not significantly different (within $\sim$ 10\%), but the line widths of the clumps become up to $3\times$ wider for TDM observations. For FDM observation, where the frequency resolution is much finer than the TDM mode used for continuum studies, the width of the clump does not reach $\approx 100$\,\kms.

\begin{table*}
    \caption{\sc The result of fitting of C and Q.}
    \centering
    \begin{threeparttable}
    \begin{tabular}{c|ccccccccccc}\hline
        FWHM (chans)\tnote{(1)}~~ & 2 & 4 & 6 & 8 & 10 & 12 & 14 & 16 & 18 & 20 & 22 \\ \hline \hline
        $a_{\rm 1}$ ($\sigma$)\tnote{(2)}~ & 6.50 & 6.42 & 6.50 & 6.50 & 6.50 & 6.50 & 6.18 & 5.87 & 5.67 & 5.54 & 4.94 \\ \hline
        $b_{\rm 1}$ \tnote{(3)}~ & 0.05 & 0.03 & 0.07 & 0.08 & 0.07 & 0.08 & 0.59 & 0.12 & 0.63 & 0.53 & 2.43 \\ \hline
        $a_{\rm 2}$ ($\sigma$)\tnote{(4)}~ & 4.71 & 3.61 & 3.03 & 2.96 & 2.40 & 2.15 & 2.16 & 1.98 & 1.00 & 2.35 & 1.71 \\ \hline
        $b_{\rm 2}$ \tnote{(5)}~ & 1.83 & 2.05 & 1.06 & 1.19 & 0.75 & 0.74 & 0.61 & 0.22 & 0.22 & 0.67 & 0.77 \\ \hline
    \end{tabular}
    \begin{tablenotes}
    \item[(1)] FWHM is shown in a unit of channel, where the typical channel width of ADF22 data is $\sim$ 18 \kms.
    \item[(2),(3)] Parameters of the function of contamination rate $Q$.
    \item[(4),(5)] Parameters of the function of completeness $C$.
    \end{tablenotes}
    \end{threeparttable}
    \label{tab:fitting}
\end{table*}

\begin{figure}
	\begin{center}
	\includegraphics[trim=0 0 0 0, width=80mm]{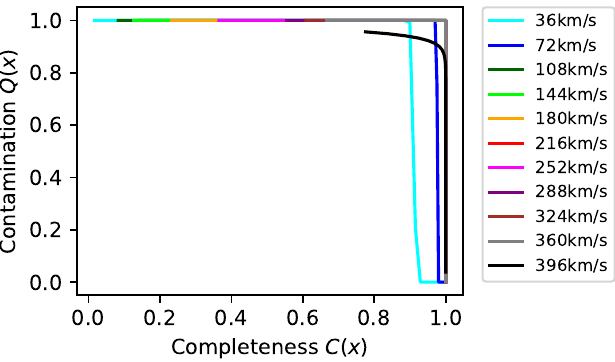}
  	\caption{The summary of the results of contamination and completeness analysis. We use this result to discuss the detectability.}
	\label{fig:contami_comp}
	\end{center}
\end{figure}

\begin{table*}
\caption{\sc The result of completeness and contamination check.}
    \begin{threeparttable}
    \scriptsize
     \begin{tabular}{c|ccccccccccc}\hline
            \multicolumn{1}{c|}{FWHM\tnote{(1)} } & \multicolumn{11}{c}{peak SN ratio [$\sigma$]}\\
            \multicolumn{1}{c|}{[chan]}&2.0&2.5&3.0&3.5&4.0&4.5&5.0&5.5&6.0&6.5&7.0\\ \hline \hline
            \multirow{4}{*}{2}&-&-&-&-&-&-&(1.0)&(1.0)&(1.0)&(0.5)&0.0\\
            &-&-&-&-&-&-&($\pm$ 0.058)&($\pm$ 0.062)&($\pm$ 0.034)&($\pm$ 0.015)&$\pm$ 0.0062\\
            &-&-&-&-&-&-&(0.65)&(0.78)&(0.88)&(0.94)&0.97\\
            &-&-&-&-&-&-&($\pm$ 0.16)&($\pm$ 0.15)&($\pm$ 0.15)&($\pm$ 0.14)&$\pm$ 0.13\\ \hline
            \multirow{4}{*}{4}&-&-&-&-&(1.0)&(1.0)&(1.0)&(1.0)&(1.0)&0.00094&0.0\\
            &-&-&-&-&($\pm$ 0.067)&($\pm$ 0.067)&($\pm$ 0.069)&($\pm$ 0.063)&($\pm$ 0.028)&$\pm$ 0.012&$\pm$ 0.0058\\
            &-&-&-&-&(0.68)&(0.8)&(0.88)&(0.93)&(0.97)&0.99&0.99\\
            &-&-&-&-&($\pm$ 0.29)&($\pm$ 0.3)&($\pm$ 0.25)&($\pm$ 0.2)&($\pm$ 0.18)&$\pm$ 0.12&$\pm$ 0.04\\ \hline
            \multirow{4}{*}{6}&-&-&-&(1.0)&(1.0)&(1.0)&(1.0)&(1.0)&(1.0)&0.49&0.0\\
            &-&-&-&($\pm$ 0.079)&($\pm$ 0.079)&($\pm$ 0.079)&($\pm$ 0.081)&($\pm$ 0.057)&($\pm$ 0.028)&$\pm$ 0.012&$\pm$ 0.0057\\
            &-&-&-&(0.83)&(0.95)&(0.99)&(1.0)&(1.0)&(1.0)&1.0&1.0\\
            &-&-&-&($\pm$ 0.14)&($\pm$ 0.14)&($\pm$ 0.089)&($\pm$ 0.072)&($\pm$ 0.041)&($\pm$ 0.042)&$\pm$ 0.026&$\pm$ 0.021\\ \hline
            \multirow{4}{*}{8}&-&-&(1.0)&(1.0)&(1.0)&(1.0)&(1.0)&(1.0)&(1.0)&(0.51)&0.0\\
            &-&-&($\pm$ 0.087)&($\pm$ 0.087)&($\pm$ 0.087)&($\pm$ 0.087)&($\pm$ 0.09)&($\pm$ 0.054)&($\pm$ 0.03)&($\pm$ 0.014)&$\pm$ 0.0061\\
            &-&-&(0.63)&(0.82)&(0.94)&(0.98)&(1.0)&(1.0)&(1.0)&(1.0)&1.0\\
            &-&-&($\pm$ 0.0)&($\pm$ 0.0)&($\pm$ 0.0)&($\pm$ 0.0)&($\pm$ 0.0)&($\pm$ 0.0)&($\pm$ 0.0)&($\pm$ 0.0)&$\pm$ 0.0\\ \hline
            \multirow{4}{*}{10}&-&(1.0)&(1.0)&(1.0)&(1.0)&(1.0)&(1.0)&(1.0)&(1.0)&(1.0)&0.43\\
            &-&($\pm$ 0.093)&($\pm$ 0.093)&($\pm$ 0.093)&($\pm$ 0.093)&($\pm$ 0.093)&($\pm$ 0.093)&($\pm$ 0.056)&($\pm$ 0.029)&($\pm$ 0.014)&$\pm$ 0.0064\\
            &-&(0.75)&(0.95)&(0.99)&(1.0)&(1.0)&(1.0)&(1.0)&(1.0)&(1.0)&1.0\\
            &-&($\pm$ 0.15)&($\pm$ 0.11)&($\pm$ 0.081)&($\pm$ 0.05)&($\pm$ 0.012)&($\pm$ 0.0)&($\pm$ 0.0)&($\pm$ 0.0)&($\pm$ 0.0)&$\pm$ 0.0\\ \hline
            \multirow{4}{*}{12}&-&(1.0)&(1.0)&(1.0)&(1.0)&(1.0)&(1.0)&(1.0)&(1.0)&(0.51)&0.0\\
            &-&($\pm$ 0.1)&($\pm$ 0.1)&($\pm$ 0.1)&($\pm$ 0.1)&($\pm$ 0.1)&($\pm$ 0.1)&($\pm$ 0.1)&($\pm$ 0.061)&($\pm$ 0.029)&$\pm$ 0.015\\
            &-&(0.87)&(0.98)&(1.0)&(1.0)&(1.0)&(1.0)&(1.0)&(1.0)&(1.0)&1.0\\
            &-&($\pm$ 0.0)&($\pm$ 0.0)&($\pm$ 0.0)&($\pm$ 0.0)&($\pm$ 0.0)&($\pm$ 0.0)&($\pm$ 0.0)&($\pm$ 0.0)&($\pm$ 0.0)&$\pm$ 0.0\\ \hline
            \multirow{4}{*}{14}&-&(1.0)&(1.0)&(1.0)&(1.0)&(1.0)&(1.0)&(0.95)&(0.67)&0.22&0.026\\
            &-&($\pm$ 0.11)&($\pm$ 0.11)&($\pm$ 0.11)&($\pm$ 0.11)&($\pm$ 0.11)&($\pm$ 0.11)&($\pm$ 0.064)&($\pm$ 0.028)&$\pm$ 0.014&$\pm$ 0.006\\
            &-&(0.92)&(0.99)&(1.0)&(1.0)&(1.0)&(1.0)&(1.0)&(1.0)&1.0&1.0\\
            &-&($\pm$ 0.097)&($\pm$ 0.053)&($\pm$ 0.0)&($\pm$ 0.0)&($\pm$ 0.0)&($\pm$ 0.0)&($\pm$ 0.0)&($\pm$ 0.0)&$\pm$ 0.0&$\pm$ 0.0\\ \hline
            \multirow{4}{*}{16}&(1.0)&(1.0)&(1.0)&(1.0)&(1.0)&(1.0)&(1.0)&(1.0)&0.074&3.1e-13&0.0\\
            &($\pm$ 0.11)&($\pm$ 0.11)&($\pm$ 0.11)&($\pm$ 0.11)&($\pm$ 0.11)&($\pm$ 0.11)&($\pm$ 0.11)&($\pm$ 0.066)&$\pm$ 0.029&$\pm$ 0.014&$\pm$ 0.0059\\
            &(0.96)&(1.0)&(1.0)&(1.0)&(1.0)&(1.0)&(1.0)&(1.0)&1.0&1.0&1.0\\
            &($\pm$ 0.12)&($\pm$ 0.0)&($\pm$ 0.0)&($\pm$ 0.0)&($\pm$ 0.0)&($\pm$ 0.0)&($\pm$ 0.0)&($\pm$ 0.0)&$\pm$ 0.0&$\pm$ 0.0&$\pm$ 0.0\\ \hline
            \multirow{4}{*}{18}&(1.0)&(1.0)&(1.0)&(1.0)&(1.0)&(1.0)&(0.93)&(0.65)&0.23&0.032&0.0015\\
            &($\pm$ 0.12)&($\pm$ 0.12)&($\pm$ 0.12)&($\pm$ 0.12)&($\pm$ 0.12)&($\pm$ 0.12)&($\pm$ 0.11)&($\pm$ 0.069)&$\pm$ 0.03&$\pm$ 0.015&$\pm$ 0.0068\\
            &(0.97)&(1.0)&(1.0)&(1.0)&(1.0)&(1.0)&(1.0)&(1.0)&1.0&1.0&1.0\\
            &($\pm$ 0.1)&($\pm$ 0.0)&($\pm$ 0.0)&($\pm$ 0.0)&($\pm$ 0.0)&($\pm$ 0.0)&($\pm$ 0.0)&($\pm$ 0.0)&$\pm$ 0.0&$\pm$ 0.0&$\pm$ 0.0\\ \hline
            \multirow{4}{*}{20}&(1.0)&(1.0)&(1.0)&(1.0)&(1.0)&(1.0)&(0.92)&(0.54)&0.11&0.0054&5.4e-05\\
            &($\pm$ 0.11)&($\pm$ 0.11)&($\pm$ 0.11)&($\pm$ 0.11)&($\pm$ 0.11)&($\pm$ 0.11)&($\pm$ 0.11)&($\pm$ 0.067)&$\pm$ 0.03&$\pm$ 0.015&$\pm$ 0.0069\\
            &(0.83)&(0.98)&(1.0)&(1.0)&(1.0)&(1.0)&(1.0)&(1.0)&1.0&1.0&1.0\\
            &($\pm$ 0.05)&($\pm$ 0.0)&($\pm$ 0.0)&($\pm$ 0.0)&($\pm$ 0.0)&($\pm$ 0.0)&($\pm$ 0.0)&($\pm$ 0.0)&$\pm$ 0.0&$\pm$ 0.0&$\pm$ 0.0\\ \hline
            \multirow{4}{*}{22}&(0.96)&(0.92)&(0.87)&(0.8)&(0.71)&(0.6)&0.49&0.37&0.27&0.18&0.12\\
            &($\pm$ 0.12)&($\pm$ 0.12)&($\pm$ 0.12)&($\pm$ 0.12)&($\pm$ 0.12)&($\pm$ 0.12)&$\pm$ 0.12&$\pm$ 0.07&$\pm$ 0.032&$\pm$ 0.017&$\pm$ 0.0072\\
            &(0.89)&(0.98)&(1.0)&(1.0)&(1.0)&(1.0)&1.0&1.0&1.0&1.0&1.0\\
            &($\pm$ 0.061)&($\pm$ 0.0)&($\pm$ 0.0)&($\pm$ 0.0)&($\pm$ 0.0)&($\pm$ 0.0)&$\pm$ 0.0&$\pm$ 0.0&$\pm$ 0.0&$\pm$ 0.0&$\pm$ 0.0\\ \hline
    \end{tabular}
    \begin{tablenotes}
    \footnotesize
    \item[(1)] FWHM is shown in a unit of channel. The typical channel width of ADF22 data is $\sim$ 18 \kms. \\
    The first, second, third, and fourth rows in each box show contamination rate $Q$, 1$~\sigma$ variance of $Q$, completeness $C$, and 1$~\sigma$ variance of $C$, respectively. The dash mark and bracket represent Category III and II, respectively. The values without brackets express the parameter classified as Category I.
    
    \end{tablenotes}
    \end{threeparttable}
\end{table*}

Therefore, we must be careful when searching for line emitters in TDM ALMA data, since the clump-like structure in the datacube would increase the contamination rate.

\subsubsection{Statistical Fluctuation}
We generate 100 mock mosaic data sets by using the replacement process. In Fig.~\ref{fig:pixeldist_res3}, we show the pixel distribution for the 12-channel smoothed SN cubes. The maximum and minimum SN values extend $\pm$ 6-8$~\sigma$, implying high-SN pixels are not unusual in smoothed mosaic data.

We search for noise clumps in the blank mock data using {\sc clumpfind} \citep{williams1994} with the same parameter set as the previous search. We calculate the number of clumps which satisfy the condition of the detection, i.e., above 6$~\sigma$ and an excess from the negative maximum for each datacube. We find the number of `detected' clumps is 0.43 $\pm$ 0.67 per datacube. Thus we expect 0-1 false-detection with $>$ 6$~\sigma$ per datacube, which is consistent with our result of false detections.

From the results above, the false detections of the \cplus\ emitter candidates can be explained by statistical fluctuation caused by the large number of the correlated pixels in the ALMA data and also the smoothing process. This is unavoidable to extract faint clumps in the interferometric data. It is therefore essential to calculate the contamination rate with a large number of mock mosaic data sets.

Note that fitting the number count of the clumps could be a useful test to ensure the Gaussian distribution of the noise. The resulting values of the mean and standard deviation of this fit should ideally be 0 and 1, respectively. If the standard deviation of this distribution is not one, it might indicate the presence of additional factors affecting the significance levels, potentially rendering the 6\,$\sigma$ detections not the expected statistical significance. However, it is important to note that the {\sc clumpfind} method is not capable of identifying clumps with low SN ratios; therefore, we did not perform this particular test. In \citet{hayatsu2017}, Fig.\,4 illustrates the number count of both negative and positive clumps above the detection limit of 4.5\,$\sigma$.

\subsection{The detectability}
We obtain $N_{\rm det}$ and $N_{\rm mock}$ from the result of the search using real and blank mock data, respectively. Also, we inject artificial sources into the blank mock single field and obtain $N_{\rm inj}$ and $N_{\rm out}$. By using these values, we calculate contamination rate ($Q$) and completeness ($C$) and fit the results with the functions in Eqs.\,4 and 5. In Tab.\,\ref{tab:fitting}, we show the parameters as the result of the fitting. In our method, the $a_2$ values are always below $a_1$ in the investigated parameter range. Therefore, most of the sources in the ADF22 field are contaminated with noise. As for the completeness, the smaller velocity width has the larger inflection value, indicating incompleteness of clumps with small velocity width. In Fig.\,\ref{fig:contamination} and \ref{fig:completeness}, we show the result of fitting for $Q(x)$ and $C(x)$, respectively and the combined result of $Q$ and $C$ is shown in Fig\,\ref{fig:contami_comp} and Tab.\,3. In the table, we also show the classification of the detectability.

\begin{figure}
	\begin{center}
	\includegraphics[trim=0 0 0 0, width=80mm]{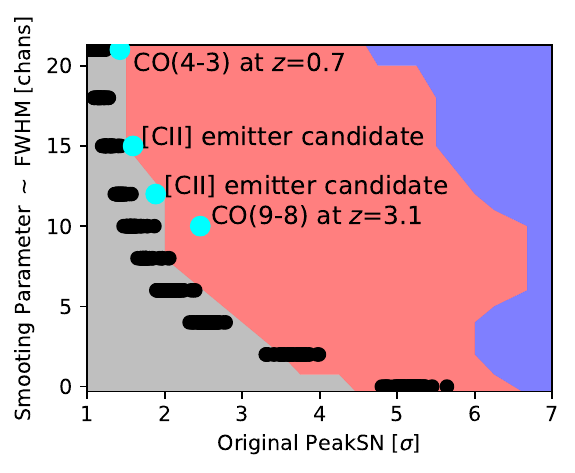}
      \caption{The result of the detection in the FWHM--peak flux plane. The black and cyan points show the detection using real ADF22 data and background show the categories derived by the contamination check and completeness analysis. The blue, red, and grey-shaded regions represent Category I, II, and III, respectively. The boundary between Category I and II are determined by $a_1$, and the boundary between II and III are determined by $a_2$. The detected \cplus\ candidates and CO emitters reside around the boundary of Category II and III, which means the candidates cannot satisfy the criteria to be treated as detections with the new method.}
	\label{fig:previousdetec}
	\end{center}
\end{figure}

In Fig.\,\ref{fig:previousdetec}, we plot the result of the search using the real data and classification in the peak SN--FWHM plane. According to our analysis, the detected sources are at the boundary between Category II and III, meaning the detections are likely affected by noise contamination. Also, for all the clumps classified as Category II, the contamination rate and completeness are both $\sim$ 1. Therefore, the corrected detection number is $\sim$ 0, which is consistent with the result of non-detection of \cplus\ emitters. We conclude that our method works to detect faint FIR emitter candidates blindly, however, the candidates are often contaminated with noise.

\section{Discussions}

\begin{table}
    \caption{The output/input ratios of the injected sources.}
    \centering
    \label{tab:inout}
    \begin{threeparttable}
    \begin{tabular}{c|cccc}\hline
            \multicolumn{1}{c|}{FWHM\tnote{(1)}~~} & \multicolumn{4}{c}{Peak S/N ratio [$\sigma$]\tnote{(2)}~~}\\
            \multicolumn{1}{c|}{[chan]}& 2.0&3.0&5.0&7.0\\ \hline \hline
            \multicolumn{1}{c|}{2} & 0.07/2.0\tnote{(3)}~~ & 0.07/1.8 & 0.10/1.6 & 0.30/1.4 \\
            \multicolumn{1}{c|}{6} & 0.17/2.1 & 0.36/1.9 & 0.37/1.7 & 0.60/1.5 \\
            \multicolumn{1}{c|}{10} & 0.17/2.2 & 0.3/2.0 & 0.65/1.8 & 0.70/1.6 \\
            \multicolumn{1}{c|}{14} & 0.20/2.3 & 0.40/2.1 & 0.60/1.9 & 0.80/1.7 \\
            \multicolumn{1}{c|}{18} & 0.30/2.4 & 0.52/2.2 & 0.65/2.0 & 0.80/1.8 \\
            \multicolumn{1}{c|}{22} & 0.41/2.5 & 0.46/2.3 & 0.67/2.1 & 0.76/1.9\\
            \hline
        \end{tabular}
    \begin{tablenotes}
    \item[(1)] Input FWHM value in units of the number of channels. The channel width is 18 \kms.
    \item[(2)] Input peak S/N ratio.
    \item[(3)] The output/input ratio of FWHM and peak S/N ratio, respectively.
    \end{tablenotes}
    \end{threeparttable}
\end{table}

\subsection{Preparation for Abundant Detection}
By using realistic mock observations, we can examine to what extent the observed LF (output of our analysis) differs from the underlying true LF. Such estimates are necessary to perform reconstruction using future, high-sensitivity observations. Note that the output/input ratio depends crucially on the luminosity considered. Essentially, we consider the contamination $Q$ and completeness $C$ to interpret or convert the output number counts by a multiplication factor $Q/(1-C)$. 

Here we discuss the output LF after correcting the contamination rate, i.e., with the `incompleteness' of the detection, since $Q$ is close to unity in a wide parameter range of interest. For simplicity, we assume that all the sources in the datacube are \cplus\ emitters and no other emission lines are detected. The impact of contamination of other line emissions can be studied by, for example, modeling the luminosity function and source properties of them.

As a simple test, we assume a Schechter function for an input \cplus\ LF; 
\begin{equation}
\phi(L) {\rm d}L = \phi^{*} \left(\frac{L}{L^{*}}\right)^{\alpha} {\rm exp}\left( - \frac{L}{L^{*}} \right) \frac{{\rm d}L}{L^{*}},
\end{equation}
where $\alpha$ = 2.36, $L^{*}$ = 2.173 $\times$10$^8$ \Lsun, and $\phi^{*}$ = 0.003 Mpc$^{-3}$, by referring to the values at local \cplus\ LF from \citet{hemmati2017}
\footnote{Note that the original function in \citet{hemmati2017} is a double power law.}. 

We generate 1000 luminosity values (sources) obeying the distribution and assign the input FWHM values from the observed correlation between $L_{\rm [CII]}$-FWHM \citep{kohandel2019}. Then we calculate the input peak flux values using Eq.\,(6), assuming the mean redshift to be 6.2 and the observed frequency as 260 GHz. We convert the peak flux to the peak SN ratio by dividing the typical RMS value of 0.8 mJy beam$^{-1}$. Then we obtain the values of completeness for the input sources for each parameter set of (FWHM, peak SN). We also obtain the output luminosity values by referring to the output/input ratio for each parameter of FWHM and peak SN. We repeat the process for 100 realizations and calculate the variance.

In order to calculate the output luminosity, we measure the output peak S/N and FWHM values of the injected sources. We fit the output line emission, assuming a single Gaussian profile, and obtain the peak flux and FWHM. We do not consider the shift of the central frequency, since it does not change the average value of the output LF. In Tab.\,~\ref{tab:inout}, we show the mean values of the output/input ratio for a set of cases. The ratio for the FWHM becomes less than unity because the edge of a line can be buried by noise. The ratio for the peak S/N ratio becomes more than unity, simply because the source with higher flux value is easier to detect. Since the output luminosity is calculated by multiplying the values, we expect the output LF would not largely shift on the horizontal axis, but the luminosity range may be broadened.

\begin{figure}
	\begin{center}
	\includegraphics[trim=0 0 0 0, width=80mm]{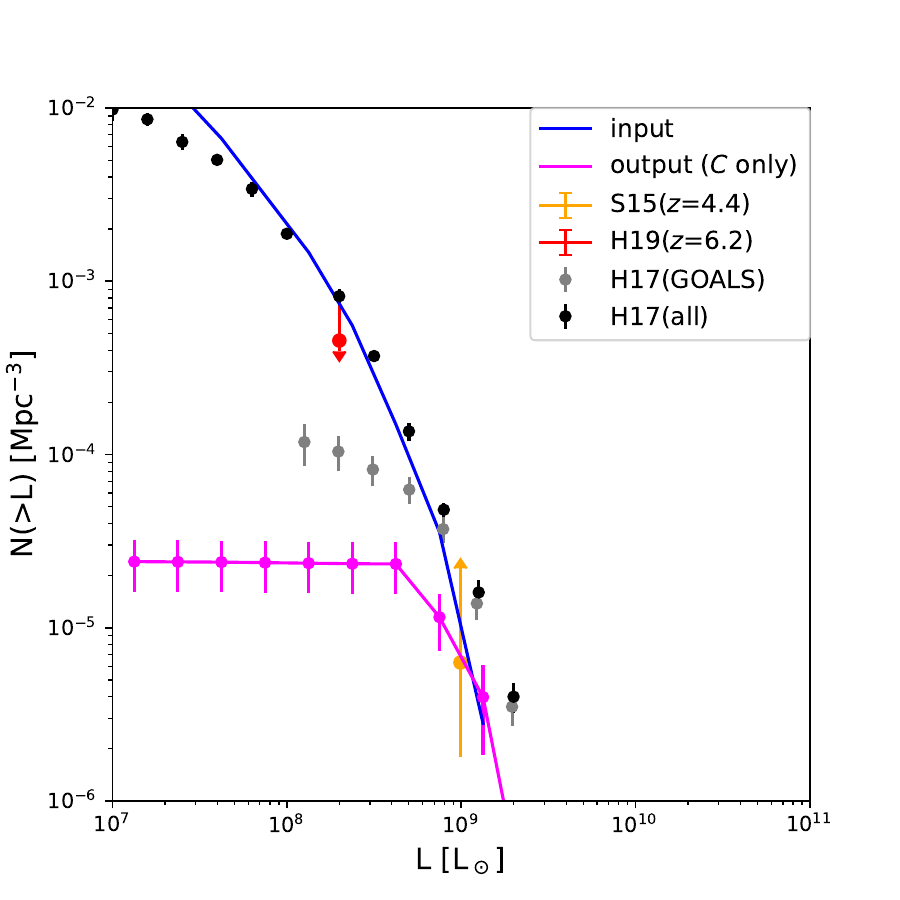}
  	\caption{The input and output \cplus\ LFs. Blue and magenta lines show the input model distribution and the output number counts with errors, respectively. We also plot a few available observational constraints in the local Universe and at high-$z$ \citep{hemmati2017, swinbank2012, hayatsu2019}. We assume a Schechter function and an empirical FWHM--peak flux distribution when injecting sources. We apply our reconstruction method for each source and multiply by the output/input ratio for the source properties. The gap between the input and output LF implies that we miss $\sim90$\% of the sources in the luminosity range of $\sim$ 5 $\times 10^8$ \Lsun, if observations have a similar sensitivity to the ADF22 but for a sufficiently large survey volume.}
	\label{fig:inout}
	\end{center}
\end{figure}

As shown in Fig.~\ref{fig:inout}, significant correction to the 'raw' number counts is required, by over an order of magnitude at $\geq$ 5 $\times 10^8$ \Lsun. The large difference between the input and output LFs is caused largely by non-detection of injected sources of category~III. If the number of detections increases in future high-sensitivity surveys, our recovery process would enable us to estimate the intrinsic LF from the measured LF (which we cannot do for normal luminosity galaxies with the current ALMA sensitivity). Also, the gap between the input and output LFs implies that we miss $\sim$ 90\% of the sources in the luminosity range of $\geq$ 5 $\times 10^8$ \\ if observations have a similar sensitivity to ADF22 but for a large enough survey volume.

\subsection{Implications for Cosmic Star-Formation History}
Our basic motivation is to understand the evolution of galaxy populations by probing the cosmic star-formation history (CSFH). In order to estimate star-formation rate density (SFRD), we derive upper limits of SFR from the non-detection of \cplus\ emitter candidates ($L_{\rm [C\,II]}$ $<$ 2\,$\times$\,10$^8$\Lsun) at $z$ = 6.2. We use the SFR--$L_{\rm [C\,II]}$ relations from \citet{delooze2014} which is calibrated using observations of nearby low-metallicity dwarf galaxies and starburst galaxies:
     \begin{eqnarray}     
    \frac{\rm SFR_{\rm [C\,II]}}{\rm M_\odot yr^{-1}} &=&  10^{-5.73 \pm 0.32}
        \left ( \frac{L_{\rm [C\,II]}}{\rm L_\odot}\right )^{0.80 \pm 0.05} {\rm (metal\mathchar"712D poor~local~dwarf),} \nonumber  \\
        \\
	\frac{\rm SFR_{\rm [C\,II]}}{\rm M_\odot yr^{-1}} &=&  10^{-7.06 \pm 0.33} 
		\left ( \frac{L_{\rm [C\,II]}}{\rm L_\odot}\right )^{1.00 \pm 0.04} {\rm (starburst).}
    \end{eqnarray}
    The SFR$_{\rm [C\,II]}$ calculated with the correlation for metal-poor local dwarfs is consistent with existing upper limits on the SFR$_{\rm UV + FIR}$. Note that the SFR--$L_{\rm [C\,II]}$ relation calibrated by observations of high-redshift galaxies is considered to be applicable to bright \cplus\ emitters with $>~10^9 \rm L_\odot$ \citep{delooze2014}, and hence it may not be accurate for normal (less luminous) \cplus\ emitters.
    The derived upper limits of the first SFRD$_{\rm [C\,II]}$ become $<$ 0.02 and 0.04 M$_\odot$ yr$^{-1}$ Mpc$^{-3}$ with the correlations for metal-poor local dwarfs and starbursts, respectively. Similarly, the 3$\sigma$ upper limits, effective survey volumes, and thus-calculated luminosity densities from the non-detection of [O\,{\sc i}] 145\,$\rm\mu$m, [N\,{\sc ii}] 122\,$\rm \mu$m, and [O\,{\sc iii}] 88\,$\mu$m lines, can be used to derive the independent SFRD at $z$ = 7, 8, and 12. The SFRs are calculated from line luminosities using the empirical relations for ultra-luminous infrared galaxies estimated by \citet{farrah2013}. Unfortunately, the SFRD constraints using the other FIR lines are still weak and thus further motivate systematic blind FIR line surveys to estimate the SFRDs at multiple redshifts simultaneously (Fig.\,17).

\begin{figure*}
	\begin{center}
	\includegraphics[width=160mm, trim= 0 0 0 0]{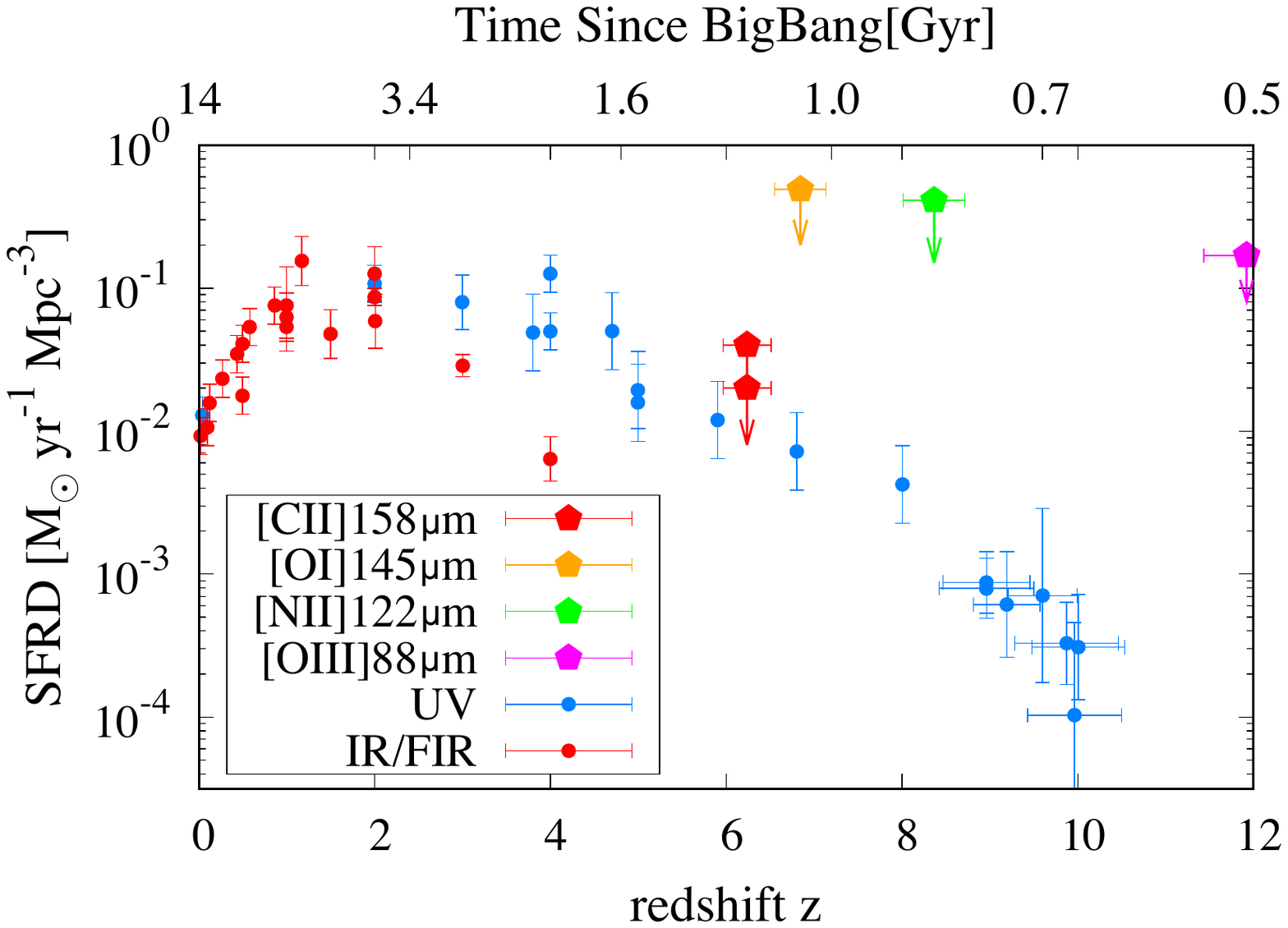}
	\end{center}
	\caption{The cosmic star-formation rate density (SFRD) as a function of redshift. We show an upper limit on the \cplus\ SFRD at $z = 6$, derived from our constraints on the \cplus\ luminosity density using SFR-L$_{\rm [CII]}$ empirical laws for metal-poor local dwarf (downward red pentagon) and starburst galaxy (upward). We also estimate upper limits for the non-detected fine-structure lines at $z = 7$--12. We plot the SFRD estimated from rest-frame ultraviolet (UV) observations \citep{oesch2013, capak2015} and in other wavebands \citep{behroozi2013}.}
\end{figure*}

\section{Summary and Conclusion}
In the previous Cycle 2 survey, two high-$z$ emitter candidates were blindly detected at a 1.1 mm wavelength, corresponding to the \cplus\ redshift of $6.0$ and $6.5$. We see the spectral feature even in the divided data in two epochs or separated data in a single field or correlation. We calculate the contamination rate by considering the negative clumps in different SPWs and different smoothing lengths from the detections of the candidates. The contamination rate calculated by using four datacubes suggested that one of the two high-$z$ emitter candidates should be robust.

The follow-up observation in ALMA Cycle 5 showed non-detections for both candidates. We have given the following speculations for the reason of the false detections in previous observations: (1) statistical fluctuation in the large mosaic data, (2) non-Gaussian noise in interferometric data, (3) spectral smoothing process, including hanning smoothing process in the visibility plane, (4) technical problems. The possibilities (1)--(3) can be evaluated by contamination rate and completeness analysis and (4) can be examined with QA3 process.

We confirm the effect of non-Gaussian noise is negligible for the ADF22 datacube by comparing the pixel distributions between real data and mock data. We compare the results of clump-finding between the time division mode (TDM) and frequency division mode (FDM) correlator datacubes and confirm that the FWHM of the clumps for the TDM case became up to $\times$ 3 wider than the FDM case. We show 0.43 $\pm$ 0.67 false-detection per datacube with the previous source-finding method using 100 blank mock-mosaic data. Thus, the underestimation of the contamination rate in the previous work is caused by fewer datacubes using four real ADF22 datacubes.

Since the QA3 process found no problem for both Cycle 2 and Cycle 5 data, we argue that false detection of high-SN ratio is unavoidable because the clump-like structure already exists in large datacubes. We also confirmed that the previously detected emitters/candidates are classified as unreliable. The LF estimation using models shows that the correction for the number count is required up to one order of magnitude at a luminosity range of $\geq 5 \times 10^8$ \Lsun. Our reconstruction method for line LF can be applied to other blind line surveys using ALMA deep field datacubes.

Altogether, observations in the mm--submm band hold promise to extract information on the star-formation activities of high-redshift galaxies.

\begin{ack}
NHH was supported by the ALMA Japan Research Grant of National Astronomical Observatory of Japan (NAOJ) Chile Observatory, NAOJ-ALMA-0071 and NAOJ-ALMA-0160, the grant of NAOJ Visiting Fellow Program supported by the Research Coordination Committee, NAOJ, and by funding from the Foundation for Promotion of Astronomy.

HU acknowledges support from JSPS KAKENHI Grant Number 20H01953.

This paper makes use of the following ALMA data: ADS/JAO.ALMA\#2013.1.00162.S, \#2013.1.00718.S, \#2017.1.00602.S., and \#2019.1.00883.S. ALMA is a partnership of ESO (representing its member states), NSF (USA) and NINS (Japan), together with NRC (Canada) and NSC and ASIAA (Taiwan) and KASI (Republic of Korea), in cooperation with the Republic of Chile. The Joint ALMA Observatory is operated by ESO, AUI/NRAO, and NAOJ.
\end{ack}

\appendix
\section{Smoothing Process}
If the signal-to-noise ratio (S/N ratio) of a clump is not large enough to confirm the detection because of significant contamination by noise, we perform spectral smoothing to increase the S/N ratio\footnote{We do not consider spatial smoothing since high-$z$ emitters should be observed as point sources.}. The questions here are what is the best smoothing parameter and to what extent we can increase the S/N ratio for line emission. For simplicity, let us consider one dimension. We assume that spectral resolution 2$\Delta v$ is small enough compared to the full-width half-maximum (FWHM) of the source, FWHM$_{\rm ori}$. We use a boxcar smoothing kernel for our line search:
\begin{equation}
 y[i+w/2] = \frac{y[i]+y[i+1]+\cdots+y[i+w]}{w},
\label{eq:boxcar}
\end{equation}
where $y[i]$ is the $i$-th arbitrary spectral channel of the data where the peak of the spectrum resides, and the integer $w$ is the smoothing width in units of the spectral channel. Note that, for simplicity, we do not apply Hanning smoothing to the CDM data in this section.

Since the spectral line emission considered here is mainly broadened by the thermal Doppler effect, we assume the line emission has a Gaussian profile. By introducing a positive constant $k$ and the velocity dispersion of the source $\sigma$, $w$ can be determined as
\begin{equation}
 w = \frac{k\sigma}{\Delta v}.
\label{eq:2_2}
\end{equation}
Note that we define the Gaussian as a function of velocity. Equation \ref{eq:2_2} expresses the relation between the smoothing parameter and the velocity width, i.e., $w \Delta v = k \sigma$.

Let the peak flux be denoted as $x_{\rm ori}$. We can express the integration of the Gaussian as follows:
\begin{equation}
 \int_{-\infty}^{\infty} x_{\rm ori} \exp\left( - \frac{(x-x_i)^2}{2\sigma^2} \right) {\rm d}v = \sqrt{2\pi} x_{\rm ori} \sigma.
\end{equation}

Since the finite integral, which is the cumulative distribution function, is expressed by the error function:
\begin{equation}
{\rm erf}(x) = \frac{2}{\sqrt{\pi}} \int_0^x e^{-t^2} {\rm d} t,
\end{equation}
thus, Eq.~\ref{eq:boxcar} is expressed using the smoothed peak flux $x_{\rm sm}$ as:
\begin{equation}
x_{\rm sm} \Delta v \sim \frac{\sqrt{2\pi} x_{\rm ori}~\sigma~{\rm erf}(k\sigma/2\sqrt{2}\sigma)}{k\sigma/\Delta v}
= \frac{\sqrt{2\pi}}{k} {\rm erf}\left( \frac{k}{2\sqrt{2}} \right) x_{\rm ori} \Delta v,
\label{eq:3.2.2}
\end{equation}

From the additivity of variance for independent distribution functions, the relation between the original RMS noise ${\rm RMS}_{\rm ori}$ and the smoothed RMS noise ${\rm RMS}_{\rm sm}$ is:
\begin{equation}
 {\rm RMS}_{\rm sm} = \frac{{\rm RMS}_{\rm ori}}{\sqrt{w}}.
\label{eq:3.2.3}
\end{equation}

Using Eq.~\ref{eq:3.2.2} and Eq.~\ref{eq:3.2.3}, we obtain the relation between the original and smoothed S/N ratio:
\begin{equation}
\frac{x_{\rm sm}}{\rm RMS_{\rm sm}} = \sqrt{\frac{2\pi\sigma}{k\Delta v}} {\rm erf}\left( \frac{k}{2\sqrt{2}} \right) \frac{x_{\rm ori}}{{\rm RMS}_{\rm ori}}
\label{eq:xsm_xori}
\end{equation}

\begin{figure}
    \begin{center}
    \includegraphics[trim=0 20 0 0, width=80mm]{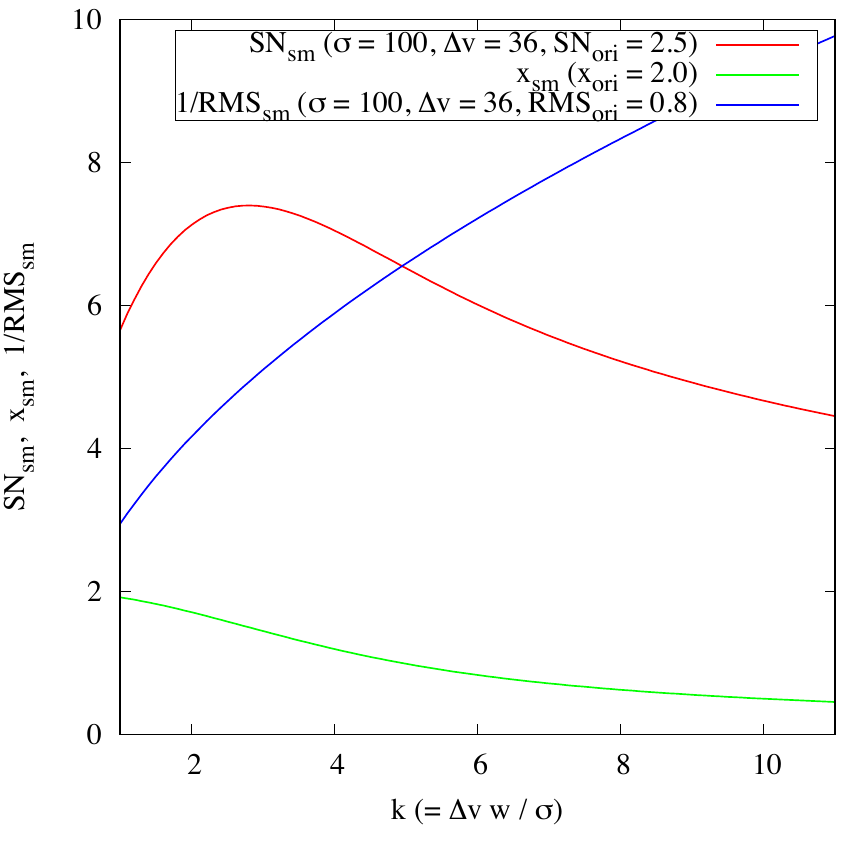}
    \end{center}
    \caption{The result of a test calculation of S/N ratio, peak flux, inverted RMS noise level after spectral smoothing as a function of normalized channel width by the velocity dispersion of the line emission. We assume a Gaussian line profile and use Eq.~\ref{eq:xsm_xori}. The inverted RMS noise value increases with the smoothing process, and the peak flux decreases. The multiplied value, $SN_{\rm sm}$, peaks when the smoothing width is close to the FWHM of the line emission.}
    \label{fig:smoothing}
\end{figure}

Fig.\,\ref{fig:smoothing} shows the calculated smoothed peak flux $x_{\rm sm}$, inverted smoothed RMS value RMS$_{\rm sm}$, and smoothed S/N ratio SN$_{\rm sm}$ as a function of $k$. The SN$_{\rm sm}$ value has a peak value at $k \sim 2.3$, and by considering $w \Delta v = k \sigma$, the SN$_{\rm sm}$ becomes the maximum when the smoothing width $\sim$ 2.3$\sigma$. This is close to the FWHM value.

To obtain the highest S/N ratios for the sources, the original data is smoothed in the spectral domain. The top-hat spectral smoothing window is set to be 0, 2, 4, ..., 12, 15, 18, ..., 21 channels, with a width corresponding to $\sim$18 km s$^{-1}$. Therefore, we search for emitters with 36--400 km s$^{-1}$ velocity widths. The upper limit is sufficient to detect normal \cplus\ emitters (e.g., \cite{kohandel2019}). Note that some dusty star-forming galaxies have a large velocity width $>$ 1000 \kms\ (e.g., \cite{venemans2012}), but we expect such galaxies can be discovered with a different method, e.g., continuum search. We use the spectral smoothing function ``boxcar'', and the velocity sampling of the output data is kept constant.

\section{Original Data}

\subsection{Cycle-2 data}
ADF22 was observed with 9 execution blocks (EBs) \footnote{In \citet{umehata2017}, dubbed as scheduling blocks.} in June 2014 and April 2015 (Proposal ID 2013.1.00162.S, PI: H. Umehata), as detailed in \citet{umehata2017}. The field was observed using Band 6 time division mode (TDM) correlator, yielding four 1.875-GHz spectral windows (SPWs) with central frequencies of 254, 256, 270, and 272\,GHz, where no significant atmosphere absorption is seen \citep{hayatsu2017}. The redshift coverage of \cplus\ line emission is $z$ = 6.0--6.5.

\begin{figure*}
	\begin{center}
	\includegraphics[trim=0 0 0 0, width=150mm]{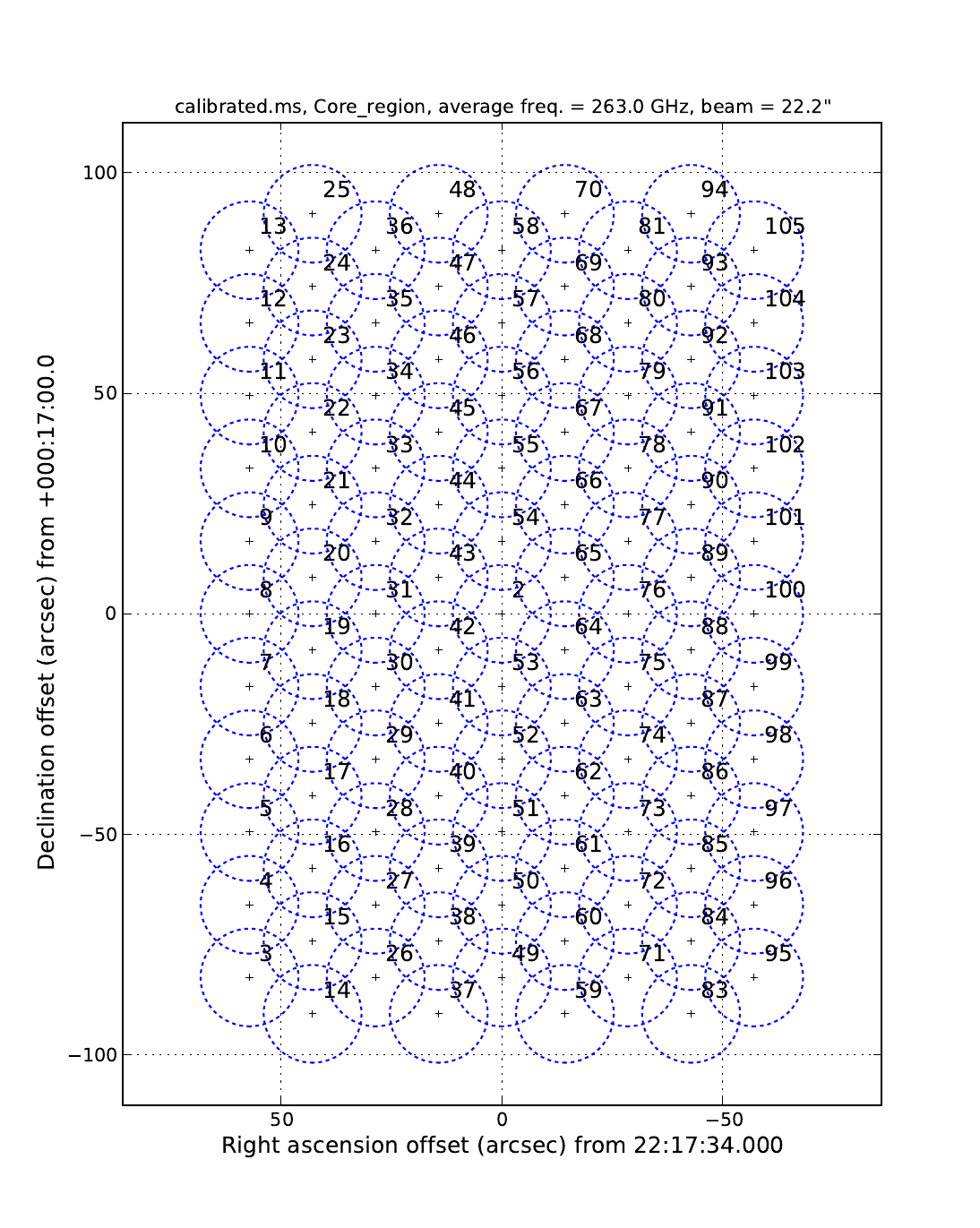}
  	\caption{The mosaic field ID. The figure is generated using \texttt{aU.plotmosaic}.}
	\label{fig:mosaicID}
	\end{center}
\end{figure*}

The field is centered on a $z$ = $3.09$ proto-cluster; RA (J2000) = 22$\rm ^h$17$\rm ^m$34$\rm^s$, Dec (J2000) = +00$^\circ$17$'$00$''$. The maximum survey field is an area of $2' \times 3'$ consisting of 103 pointing fields, but 4 EBs are 80 pointing and 1 EB is 40 pointing observation. The configurations using 33--36 12-m antennas are C34-2 and C34-4. In order to search for faint emission line sources, we use high-sensitivity data of 80 pointing fields; Field 2 - Field 81 and search in a rectangular area of $\sim$ 5 arcmin$^2$, and a frequency coverage of 253.1--272.8 GHz. Fig.\,\ref{fig:mosaicID} shows the field ID, beam response, survey area, and the position of the sources / candidates detected. The effective survey area for \cplus\ corresponds to about 29 comoving Mpc$^2$ and the survey volume is $\sim$ 2.2 $\times$ $10^3$ comoving Mpc$^3$ at $z$ = $6.2$.

The data observed in 2014 and 2015 have typical angular resolutions of $0''.54 \times 0''.49$ and $1''.24 \times 0''.87$, respectively, and the combined data has $0''.72 \times 0''.62$ angular resolution. The on-source time per pointing in the fields is 2.5 min. and 2.0 min. for 2014 and 2015, respectively. The precipitable water vapor (PWV) is 0.3--1.3\,mm. We make 'dirty' continuum-subtracted datacubes using {\sc uvcontsub} and {\sc clean}. For the selected survey area, the four SPWs have root-mean-square (RMS) noise levels of 0.7, 0.7, 0.8, or 0.9 mJy beam$^{-1}$ at a 36 km s$^{-1}$ velocity resolution.

\subsection{Cycle-5 data}
In order to confirm the blind line detections of the two \cplus\ emission-line candidates, observations were undertaken in April 2017 (2017.1.00602.S, PI: N.\,H.~Hayatsu) in array configuration C43-3, where the PWV was 1.4\,mm. On-source integration times were $\approx 11$\,min, and used frequency division mode (FDM), with $4 \times 1920$ dual-polarisation channels over a bandwidth of 7.5\,GHz, with four 1.875-GHz SPWs. Single fields centered on the targets were observed at frequencies of 251.284 and 261.183\,GHz for ADF22-LineA and ADF22-LineB, respectively.

The median noise value is $\approx 0.77$\,mJy\,beam$^{-1}$ at $\approx 36$\,km\,s$^{-1}$ spectral resolution for both datasets, similar to the Cycle-2 data; the angular resolution was around $1''.0 \times 0''.6$.

Also, Band 3 observations for one of the candidates were performed.

\section{The Previous Method}
In this section, we briefly describe the previous method to extract line emitter candidates. The additional process to ensure the data quality is also shown.

We first spectrally smooth the data to obtain high signal-to-noise (S/N) ratios. The top-hat spectral smoothing window is set to be 0, 2, 4, ..., 12, 15, 18, ..., 21 slices, with a slice width corresponding to $\sim$18 km s$^{-1}$. We use the spectral smoothing function ``boxcar'' so that the velocity sampling of the output data is kept constant. As each spectral data slice has a different RMS value as shown in Figure 1, we normalize each slice by its RMS. We call a datacube thus-generated as an ``S/N cube''.

We use {\sc clumpfind} \citep{williams1994} to search for emission line sources in the S/N cube. We search for sources with a threshold value ``low'' of {\sc clumpfind} of $\geq$ 4.5. We then do `matching' of the clumps detected at the same position between the S/N cubes in the same SPW with different resolutions and retain the clump that has the maximum S/N ratio. We select clumps that have the S/N ratio larger than $6.0 \sigma$ and also larger than the maximum negative S/N ratio measured in the inverted S/N cube in each SPW, in order to avoid contamination by spurious sources (e.g.,\,\cite{hatsukade2016}). We also check line spectral features of the detected clumps (sources) in the datacube separately for those observed in 2014 and 2015. Additionally, we visually check the non-Gaussian feature in the smoothed datacube.

For the detected sources, we search for their counterparts in optical--NIR wavebands.

The blind line search described in \citet{hayatsu2017} resulted in the detection of two emission lines, at 253.79 and 269.92\,GHz, with J2000 positions: (RA, Dec) = ($22^{\rm h} 17^{\rm m} 37.43^{\rm s}$, $+00^\circ 17' 10''.7$) and ($22^{\rm h} 17^{\rm m} 31.95^{\rm s}$, $+00^\circ 18' 20''.3$), respectively. The lines were seen in two independent subsets of the data \citep{hayatsu2017}.

\section{Cross-Check of the Detection and Non-Detection}
We plot pixel distributions of the datacube and zeroth moment maps; an integrated value of the spectrum, using {\sc immoment} and checked spectral features by binning, and dividing data into two polarizations, 'XX' and 'YY'. We also confirmed the other over-lapping pointings in the Cycle 2 data, looked separately at each of the polarizations, and visually checked weak spectral features for each dataset. The corresponding field IDs, i.e., the pointings closest to those positions, are 6, 7, and 18 for ADF22-LineA and 79, 80, and 91 for ADF22-LineB. The field of view (down to a pb response of 20\%) of each pointing is $16"$ with the pointings closest to each candidate being 18 and 80, respectively.

\begin{figure*}
	\begin{center}
	\includegraphics[trim=0 0 0 0, width=150mm]{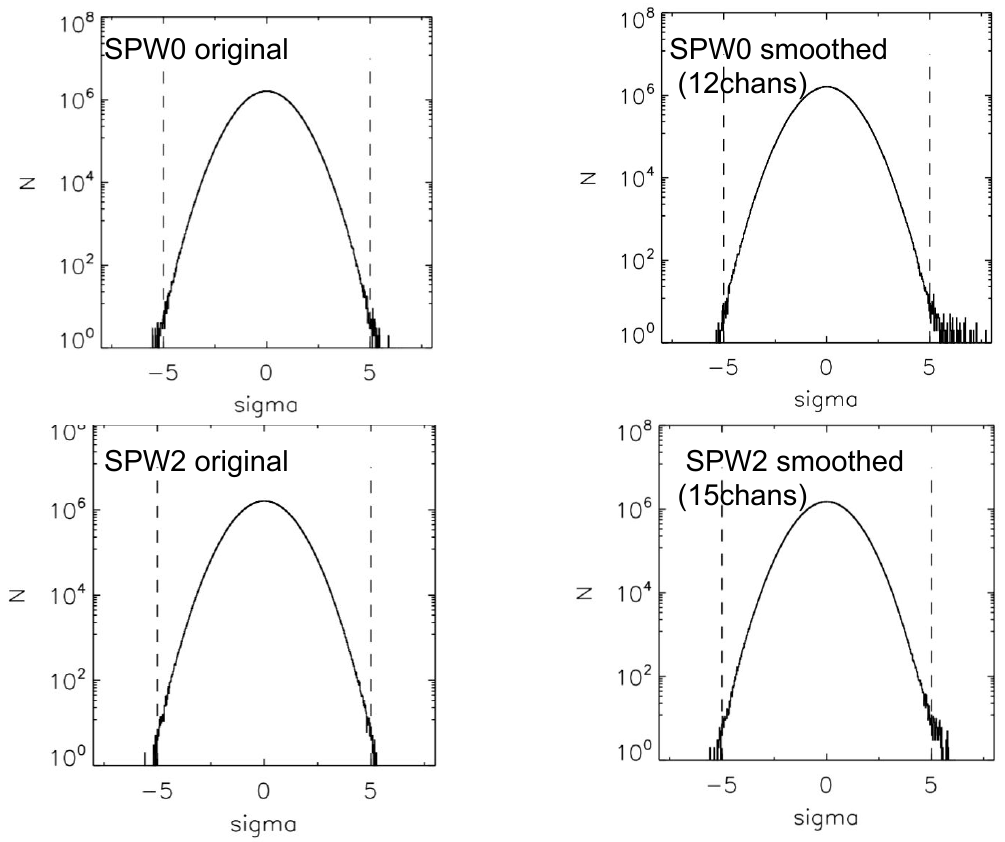}
  	\caption{The sigma distribution of the original datacube. The distributions are well fitted with Gaussian, while positive excess due to both real sources (LineC and D) and noise (LineA and B) are shown in the smoothed datacube.}
	\label{fig:oridist}
	\end{center}
\end{figure*}

Fig.\ref{fig:oridist} shows the noise distribution for the original data. The number of the pixels above 6 sigma is 5 and 78 for Spw0 datacube for original and 12 channel smoothed data, respectively, and 0 and 4 for Spw2 datacube for original and 15 channel smoothed data, respectively. All these pixels arise from the detected clumps i.e., LineA and C for Spw0 and Line B and D for Spw2. Note that there are no negative clumps below -6$\sigma$ in these Spws, while Spw3 detects one -6.3 $\sigma$ clump in the 18 channel smoothed cube.

For the Cycle 2 data, we detect the candidates as point sources, with peak S/N ratios of $\approx4 \sigma$ for Lines A and B. For the Cycle 5 data, peaks above $2\sigma$ were not detected within $1''.0$ of the positions of the candidates. Note that a known continuum source ADF22.4, is detected in the same pointing as ADF22-LineA \citep{umehata2017} for the data from Cycles 2 and 5, we checked the consistency between the positions and continuum flux densities using {\sc imfit}.

\end{document}